\documentclass[5p]{elsarticle}

\usepackage{amssymb}
\usepackage{amsmath}

\journal{Sustainable Computing: Informatics and Systems}

\usepackage{algorithm, algorithmic}
\usepackage{multirow, multicol, makecell}

\usepackage{xcolor}

\newcommand{\review}[1]{{{#1}}}
\newcommand{\rew}[1]{#1}

\PassOptionsToPackage{hyphens}{url}\usepackage{hyperref}
\begin{document}

\begin{frontmatter}



\title{Spatio-Temporal Shifting to Reduce Carbon, Water, and Land Use Footprints of Cloud Workloads}


\author{Giulio~Attenni$^{a}$} 

\author{Youssef~Moawad$^{b}$} 

\author{Novella~Bartolini$^{a}$} 

\author{Lauritz~Thamsen$^{b}$} 

\affiliation{organization={"La Sapienza" University of Rome},
            addressline={Department of Computer Science}, 
            city={Rome},
            country={Italy}}

\affiliation{organization={University of Glasgow},
            addressline={School of Computing Science}, 
            city={Glasgow},
            country={United Kingdom}}

\begin{abstract}
In this paper, we investigate the potential of spatial and temporal cloud workload shifting to reduce carbon, water, and land use footprints.
Specifically, we perform a simulation study leveraging publicly available data on the cloud infrastructure of major providers (AWS and Azure) as well as real-world workload traces (big data analytics and FaaS) and grid mix data to consider two different scenarios.
Our simulation results indicate that spatial shifting can substantially lower carbon, water, and land use footprints. 
In the FaaS applications, shifting the spatiotemporal workload achieves carbon savings of up to 85\%, water savings of around 50\%, and reductions in land use of up to 45\%, all while optimizing for the respective factors.  Mixed optimization yields results comparable to those of land use alone. For big data workloads, spatiotemporal shifting delivers reductions of up to 45\% in carbon emissions, 40\% in water consumption, and nearly 40\% in land use when optimized for the respective factors.
Temporal shifting also decreases the footprint, though to a lesser extent. When applied together, the two strategies yield the greatest overall reduction, driven mainly by spatial shifting with temporal adjustments providing an additional, incremental benefit. Sensitivity analysis  demonstrates that such shifting is robust to prediction errors in grid mix data and to variations across different seasons. 
\end{abstract}


\end{frontmatter}



\emergencystretch=1em
\sloppy

\section{Introduction}
The Information and Communication Technology (ICT) sector is estimated to be responsible for 1.8\%–2.8\% of global GHG emissions~\cite{FREITAG2021100340} and this share is expected to rise in the future~\cite{belkhir2018assessing}.
In particular, data centers have become one of the most energy-intensive infrastructures, with their electricity consumption projected to reach around 945 TWh globally by 2030~\cite{iea_dc2030}.
This growth is driven by data-intensive applications and services, including artificial intelligence and big data analytics. 

The sustainability of cloud computing has been studied, with a primary focus on reducing carbon emissions. A wide body of work investigates carbon-aware scheduling and orchestration, exploiting spatial~\cite{10.1145/2342356.2342398, 10.1145/3604930.3605711, 10.1145/3631295.3631396, 10.1145/3447555.3466582, ZHENG20202208}, temporal~\cite{10.1145/3626788, 10.1145/3464298.3493399, piontek2023carbon}, and spatio-temporal~\cite{7345588, 10.1145/3634769.3634812, 10.1145/3627703.3650079, 10305816, 10.1145/3575813.3595197} shifting of workloads to align demand with low-carbon electricity availability.

Recently, studies have started to address the significant environmental impacts of cloud computing beyond carbon emissions. In particular, the impacts of water consumption have seen attention~\cite{li2023making, jiang2025waterwise}. Additionally, large-scale land use, as associated with data center construction, can lead to habitat disruption and biodiversity loss~\cite{newbold2015global}. 
These factors underscore the need for a multi-dimensional approach to evaluating and optimizing the environmental footprint of cloud computing.
To the best of our knowledge, land use has not been systematically integrated into sustainability-aware scheduling of cloud workloads before. 
Therefore, this gap \rew{is addressed} by introducing \emph{Land Usage Effectiveness (LUE)} as a metric and incorporating it into the multi-dimensional optimization of workload placement. 

In this paper, \rew{it is investigated} the spatial and temporal shifting of cloud workloads as strategies to reduce carbon, water, and land use footprints. An extensive simulation study \rew{is conducted} using real-world data from major public cloud providers (Azure and AWS) and representative cloud workloads (FaaS and big data analytics), demonstrating the potential of multi-dimensional sustainability-aware scheduling.

In summary, our main contributions are:
\begin{itemize}
    \item A holistic formulation for workload scheduling that optimizes carbon, water, and land use impacts, leveraging temporal and spatial cloud workload shifting \rew{is proposed}

    \item A systematic evaluation based on real-world workload traces and grid mix data, quantifying trade-offs under spatial, temporal, and spatio-temporal shifting scenarios \rew{is provided}. \review{A comparison with state-of-the-art solutions is presented and a Pareto Analysis of spatial shifting is included to further analyze trade-offs.} 
    Moreover, sensitivity studies on workload flexibility, grid mix prediction errors, and seasonal variations \rew{were conducted}.

    \item All code and data from this study \rew{is released} to support transparency, reproducibility, and further research \footnote{\url{https://github.com/GlasgowC3lab/CaWaLand}}.
\end{itemize}

The rest of the paper is organized as follows. 
\review{
Section~\ref{sec:background_rel_work} provides background on data center environmental impacts and reviews related work.
}
\review{
Section~\ref{sec:methodology} presents the methodology, including the proposed system model, problem formulation and optimization algorithms.
Moreover, it
}
presents our simulation setup, including a detailed description of the scenario characteristics such as workload types, data center specifications, and grid datasets. Section~\ref{sec:results} presents the results of our empirical evaluation and discusses our findings as well as threats to validity. 
Finally, Section~\ref{sec:conclusions} concludes the paper and outlines directions for future research.

\section{Background \& Related Work}\label{sec:background_rel_work}
\review{This section presents the background necessary to understand the environmental impacts of data centers (Section~\ref{sec:background}), followed by a 
comprehensive review of the state-of-the-art in sustainability-aware cloud 
computing (Section~\ref{sec:relwork}).}
\subsection{Background - Data Center Impacts}
\label{sec:background}

Data centers rank among the world's most energy intensive facilities, consuming vast quantities of electricity to run servers, cooling infrastructure, and backup power systems. This substantial energy demand translates into considerable greenhouse gas (GHG) emissions, particularly in regions that rely heavily on fossil fuels for electricity generation. As a result, the environmental footprint of any given data center is closely linked to the composition of its local power grid.

The energy mix influences the \emph{Carbon Intensity (CI)}, which measures the greenhouse gas emissions per unit of electricity consumed (typically expressed in gCO\textsubscript{2}e/kWh); the \emph{Energy Water Intensity Factor (EWIF)}, which captures the amount of water withdrawn or consumed during electricity generation (measured in liters/kWh), which varies with the cooling technologies; and the \emph{Energy Land Intensity Factor (ELIF)}, which quantifies the land use required for producing electricity (expressed in m\textsubscript{2}/kWh).  
Example estimates of these intensities are reported in~\cite{ipcc2014_ar5_annexIII, nrel2011_water, plos2022_landuse}. 
These are also the main sources used in our experiments  (see Table~\ref{tab:intensities}), and in the discussion section, we will analyze how these coefficients relate to each other. 

The efficiency of data center operations also depends on the infrastructure and technologies used.
The \emph{Power Usage Effectiveness (PUE)} 
is defined as the ratio of the total energy consumed by a data center facility to the energy consumed by its IT equipment~\cite{pue}. A PUE of 1.0 indicates ideal efficiency, where all energy is used exclusively for computation.

Moreover, the efficiency of cooling systems, which are essential to prevent overheating of servers, can significantly affect water usage. The choice of cooling technology (e.g., air cooling vs. liquid cooling) can lead to different water consumption. The \emph{Water Usage Effectiveness (WUE)}
measures the water consumption of a data center relative to its IT energy usage, typically expressed in liters of water per kilowatt-hour (l/kWh)~\cite{wue}.

Finally, since these infrastructures occupy a significant amount of land, the \emph{Land Usage Effectiveness (LUE)}~\cite{lue}
should also be considered. To the best of our knowledge, this metric has not been standardized, and it has not been used previously.
Similar to PUE and WUE, the LUE metric aims to quantify the physical land footprint of data centers in relation to IT energy consumption, capturing the spatial resource requirements of large-scale compute infrastructure. Thus, for the scope of our work, \rew{LUE is defined} as the ratio of the total land occupied by the data center property and the IT energy consumption, and it is, therefore, expressed in square meters per kilowatt-hour (m$^2$/kWh).
\review{ADD here more on LUE?}

The carbon associated with the extraction, processing, and transportation of construction materials, as well as with the production of IT equipment (known as embodied carbon) accounts for a share of the total carbon footprint of a data center. However, in this work, only the share of operational carbon emissions of the workloads \rew{is considered}.


\subsection{Related Work}
\label{sec:relwork}

Research into cloud orchestration has generated a substantial body of work aimed at improving resource management and operational efficiency~\cite{10.1145/3054177}. Within this space, two areas have attracted considerable attention: optimizing energy consumption and incorporating carbon awareness into system design~\cite{10.1145/3464298.3493399,10.1145/3634769.3634812, 10.1145/2342356.2342398, 7345588, 10.1145/3604930.3605711, 10305816, piontek2023carbon, 10.1145/3631295.3631396, 10.1145/3626788, 10.1145/3575813.3595197}
\review{\cite{10.1145/3754598.3754673, 10749780, su17146433, 11250739, 9770383}}.

Recently, the impact on water resources of computation has also raised concerns, but few studies have focused on this issue~\cite{li2023making, jiang2025waterwise}.
Different from previous works, land use \rew{is included} in an approach that simultaneously assesses and optimizes the overall impact in terms of carbon, water, and land use footprint.
\subsubsection{Carbon-Aware Computing}
In cloud computing, carbon-aware optimization focuses on aligning workload execution with variations in the carbon intensity of the electricity supply over time and/or across different geographical areas.









\paragraph{Temporal Shifting}
In~\cite{10.1145/3464298.3493399}, the temporal shifting benefits considering different regions \rew{is analyzed}, demonstrating the carbon savings potential of delay-tolerant workloads in Germany, Great Britain, France, and California.

Hanafy et al.~\cite{10.1145/3626788} propose CarbonScaler, where cloud batch jobs dynamically adjust server allocations based on grid carbon intensity, and demonstrate the performance of this approach in a Kubernetes prototype.

Piontek et al.~\cite{piontek2023carbon} introduce a carbon-aware Kubernetes scheduler that leverages a prediction algorithm to defer non-critical workloads in time, thereby reducing overall carbon emissions.

\review{Schweisgut et al.~\cite{10.1145/3754598.3754673} study carbon-aware temporal shifting for workflows with a fixed task mapping and ordering, subject to a deadline constraint, and show that the problem is solvable in polynomial time for a single processor but becomes NP-hard for two or more processors, motivating heuristic solutions for the general case.}

\paragraph{Spatial Shifting}




Gao et al.~\cite{10.1145/2342356.2342398} present a scheduler that manages traffic distribution across data centers, balancing a three-way trade-off among access latency, electricity cost, and carbon footprint. 

Building on the idea of geo-distributed scheduling, Maji et al.~\cite{10.1145/3604930.3605711} propose a load balancing solution for VMware's Avi Global Server Load Balancer, employing a linear scoring function to select the most suitable data center based on marginal carbon intensity and client-to-data-center distance. 

Similarly, Chadha et al.~\cite{10.1145/3631295.3631396} introduce GreenCourier, a Kubernetes scheduler aimed at curbing the carbon emissions of serverless functions deployed across geographically distributed regions.

\review{
In the serverless domain specifically, Serenari et al. ~\cite{10749780} propose GreenWhisk, a carbon-aware extension of Apache OpenWhisk that balances function-placement locality against real-time marginal carbon intensity via a weighted consistent-hashing load balancer.
}

Lindberg et al.~\cite{10.1145/3447555.3466582} propose a metric, named locational marginal carbon emission, to guide spatial shifting.

Zheng et al.~\cite{ZHENG20202208} evaluate the idea of migrating workloads between data centers to reduce curtailment. It has been shown that migrating workloads from fossil-fuel-powered grids to renewable-energy-powered grids during curtailment hours simultaneously reduces emissions and absorbs excess renewable generation.

\paragraph{Spatio-Temporal Shifting}




Zhou et al.~\cite{7345588} apply Lyapunov optimization to jointly perform load balancing across geo-distributed data centers, capacity right-sizing through powering down idle servers, and server speed scaling via CPU frequency adjustment, with the goal of balancing a three-way trade-off among electricity cost, SLA requirements, and emission reduction targets.

Souza et al.~\cite{10.1145/3634769.3634812} propose a provisioner that simultaneously minimizes the number of active servers and their associated carbon emissions. 

Sukprasert et al.~\cite{10.1145/3627703.3650079} examine the benefits and limitations of carbon-aware spatiotemporal shifting for cloud workloads, offering insight into its practical applicability. 

Cordingly et al.~\cite{10305816} introduce a prototype for computing resource aggregation that reduces the carbon footprint of serverless applications through carbon-aware load distribution.

Lin et al.~\cite{10.1145/3575813.3595197} propose a coordinated scheduling framework that aligns data center capacity planning with grid dynamics in both spatial and temporal dimensions to avoid increasing carbon emissions, energy costs, and reducing grid reliability.

\review{
More recently, Asadov et al.~\cite{su17146433} propose a Kubernetes pod-level scheduler that jointly minimizes operational and embodied carbon across nodes and time slots, structurally similar to our cost formulation but restricted 
to a single (carbon) objective and without accounting for migration cost. 
}

\review{Hall et al.~\cite{11250739} propose a distributionally-robust optimization framework for joint spatial and temporal load shifting across a data center fleet, extending Google's production carbon-aware system~\cite{9770383} with formal probabilistic performance guarantees via Conditional Value-at-Risk constraints. Their approach targets carbon footprint and infrastructure cost jointly.}

\subsubsection{Beyond Carbon-Aware Computing}
Recent research has begun broadening the scope of environmental impact assessment for cloud and AI workloads beyond carbon emissions alone.

Li et al.~\cite{li2023making} shed light on the often-overlooked water footprint of AI models, revealing the substantial freshwater consumption associated with both cooling systems and electricity generation, and estimating that training large AI models can result in hundreds of thousands of liters of direct water use. 

Jiang et al.~\cite{jiang2025waterwise} propose WaterWise, a job scheduling framework that co-optimizes carbon and water sustainability across geographically distributed data centers. Their approach relies on a Mixed Integer Linear Programming (MILP)-based scheduler that dynamically routes workloads to regions with favorable carbon and water conditions, while respecting delay tolerance constraints. This line of work resonates with our own, which similarly advocates for multi-dimensional sustainability metrics in cloud workload allocation and underscores the need for comprehensive, sustainability-aware orchestration strategies.

Within prior literature,~\cite{jiang2025waterwise} is the closest to our work. While WaterWise represents an important step toward integrating water sustainability, its methodology suffers from two major limitations: it simplifies regional operational characteristics by assuming constant data center efficiency (PUE set to 1.2 for all regions), and it relies on a linear regression based on the wet-bulb temperature for estimating WUE values. As shown in our previous work\cite{loco24}, this method can lead to estimations that differ substantially from the values declared by the providers, probably because it derives from a study conducted on a specific cooling technology~\cite{TCC.2015.2453982}. Instead, \rew{our focus is} more on regional differences beyond grid intensities using both region-specific PUE and WUE values declared by providers.
Furthermore, to the best of our knowledge, land use has not previously been considered alongside carbon and water footprints in cloud workload allocation. 

\subsubsection{Comparison With Own Previous Work}

\rew{Our} idea of holistic sustainable cloud workload allocation \rew{was presented} at the 1st International Workshop on Low Carbon Computing~\cite{loco24}. 
Unlike our workshop paper, this paper focuses specifically on carbon, water and land use as highly relevant yet increasingly assessable impact factors.
Furthermore, temporal shifting is \rew{also included}, and \rew{the assumption that} workloads can be migrated live and without costs \rew{is dropped}, but instead \rew{the focus moves to} spatial shifting of workloads prior to their execution in our simulations.
Moreover, to ensure the relevance of our simulations, real-world workload traces instead of synthetic data to simulate jobs \rew{are used} in this paper.

\rew{Previous works focused also} on carbon-aware temporal load shifting~\cite{10.1145/3464298.3493399}, as well as admission control policies~\cite{10.1145/3632775.3639589} and client selection strategies~\cite{10.1007/978-3-031-12597-3_14} to leverage renewable excess energy for machine learning workloads. Similarly, results on carbon-aware scheduling and scaling of scientific workflows~\cite{11044837} \rew{have been published}. However, these works aim exclusively at reducing carbon emissions and do not consider water or land use efficiencies.

\section{Methodology}
\label{sec:methodology}
\review{This section presents the methodology adopted in this study. It begins by introducing the system model and formalizing the problem statement (Section~\ref{sec:problem}). A methodology for evaluating the carbon, water, and land use footprints associated with a job request is then defined (Section~\ref{sec:footprints}), followed by the scheduling methods considered in the evaluation (Section~\ref{sec:scheduling}). The experiment design, including the workload scenarios and their configuration, is presented in Section~\ref{sec:experiment_design}. Finally, the experimental setup, covering the hardware and software environment, data sources, and simulation 
parameters, is described in Section~\ref{sec:setup}.}

\subsection{System Model and Problem Statement}
\label{sec:problem}
Before presenting the methodology of our study, \rew{a formalization of} the problem \rew{is presented} in this section.
{
\begin{table}[]
    \centering
    
    \begin{tabular}{l|p{5cm}}
         \textit{Symbol} & \textit{Description} \\ \hline
         $t_0$, $H$, $\delta_t$, ${\cal{T}}$ & period start, end, step duration and set of steps  \\

         $d$, ${\cal{D}}$ & region/data center and set of regions/data centers  \\
         $PUE_d$ & region PUE \\
         $WUE_d$ & region WUE \\
         $LUE_d$ & region LUE \\
         $A_d$ & region land occupation area \\
         $IT_d$ & region IT energy usage \\
         ${\cal{C}}_d(t)$ & region carbon footprint \\
         ${\cal{W}}_d(t)$ & region water footprint \\
         ${\cal{L}}_d(t)$ & region land footprint \\
         $CI^e_{G_d}(t)$ & region Energy Carbon Intensity prediction \\
         $EWIF^e_{G_d}(t)$ & region Energy Water Intensity Factor prediction\\
         $ELIF^e_{G_d}(t)$ & region Energy Land Intensity Factor prediction \\
         
         $p_d(t)$ & data center footprint profile at time t \\
         $p^{norm}_d(t)$ & normalized data center footprint profile at time t \\
         $p^{norm}_{global}(t)$ & global footprint profile \\
         
         $v$, ${\cal{V}}$ & VM instance and set of VM instances  \\
         $P_{v,j}$ & VM $v$ power draw for job $j$ \\
         $B_v$ & VM's network bandwidth  \\

         $j$, ${\cal{J}}$ & job and set of jobs  \\
         $o_j$, $a_j$, $t_j$& job's arrival location and arrival and deadline time \\
         $s_j$ , ${\cal{D}}[j]$ & job's data size and region availability  \\
         $v_j$ , $n_j$ & job's configuration (VM instance and nr. of nodes)  \\
         $r_j$ & job's expected runtime  \\
        
         $C(j,d,t)$ & scheduling footprints \\
         $M(j,d,t)$ & migration/data transfer footprints \\
         $NEI$ & Network Energy Intensity \\
         $L(j,d)$ & migration/data transfer latency \\
         $E(j,d,t)$ & execution footprint
    \end{tabular}

\caption{Notations}\label{tab:notation}
\end{table}
}
\subsubsection{System Model}

First of all, let us consider the temporal horizon $H$, and let us discretize the time. Thus, let us define ${\cal{T}} = [t_0, t_0+\delta_t, t_0+2\delta_t, ..., H]$, as the set of time steps. 
Let us also define ${\cal{D}}$ as the set of regions/data centers, and consider a data center $d$ and a time $t$. To evaluate its impact accounting for multiple factors (i.e., carbon, water, and land use), its sustainability profile \rew{is defined} as follows:
\begin{equation}
    \forall t \in {\cal{T}}, \ d \in {\cal{D}} \quad \mathbf{p}_d(t) = 
    \begin{pmatrix}
        {\cal{C}}_d(t)\\
        {\cal{W}}_d(t)\\
        {\cal{L}}_d(t)\\
    \end{pmatrix}
\end{equation}
where:
\begin{itemize}
    \item ${\cal{C}}_d(t)$ is the predicted carbon intensity of data center $d$ from time $t$ to time $t+\delta_t$ and it is defined as \begin{equation}
        CI^e_{G_d}(t) \times PUE_{d}
    \end{equation} 
    It is measured in \(\frac{gCO_2}{kWh}\) and represents the carbon emitted per kWh consumed. The carbon intensity prediction of the regional grid of data center $d$  at time $t$ \rew{is considered}. Namely, $CI^e_{G_d}(t)$, where $e$ is the predictive model and $G_d$ is the grid.
    This intensity is multiplied by the Power Usage Effectiveness of the data center, $PUE_d$, to account for the overall energy usage of a data center for each kWh of computation.
    \item ${\cal{W}}_d(t)$ is the predicted water intensity and it is defined as
    \begin{equation}
        WUE_d + EWIF^e_{G_d}(t) \times PUE_{d}
    \end{equation}
    The water intensity considering both the on-site and grid contributions \rew{is computed}.
    It is measured in \(\frac{l}{kWh}\) and represents the water consumed per kWh, considering the average Energy Water Intensity Factor prediction of the regional grid, $EWIF^e_{G_d}(t)$, and the Water Usage Effectiveness of the data center, $WUE_d$. 
    \item ${\cal{L}}_d(t)$ is the predicted land use intensity and it is defined as
    \begin{equation}
        LUE_d + ELIF^e_{G_d}(t) \times PUE_{d} 
    \end{equation}
    Similarly to the water intensity, both on-site and grid contributions \rew{are considered}.
    It is measured in \(\frac{m^2}{kWh}\), and it is the amount of land used per kWh considering the Energy Land Intensity Factor prediction of the regional grid at time $t$, $ELIF^e_{G_d}(t)$, and the Land Usage Effectiveness, defined as \(LUE_d  = \frac{A_d}{P^{IT}_d}\). In this formula, \(A_d\) is the total area of land occupied by the data center property and \({P^{IT}_d}\)  is the energy used to power the IT equipment.
    \review{
    Like PUE and WUE, LUE is defined as a simple efficiency ratio relating physical resources to IT energy consumption rather than as an ecologically weighted impact indicator. This ensures that LUE remains consistent with the established family of data center efficiency metrics: all three metrics quantify a resource footprint per unit of energy consumed rather than the ecological value of that resource. However, extending LUE to weight land by its biodiversity or ecosystem value would be beyond the scope of this work. 
    }
    
\end{itemize}
The aforementioned energy intensity factors of the electric grid intensity depends on the energy mix, which is both time and space-dependent. 
Thus, let us define $S$ as the list of sources in the energy mix, namely $\{Solar$, $Wind$, $Hydro$, $Geothermal$, $Biomass$, $Nuclear$, $Coal$, $Gas$, $Oil$, $Unknown\}$, and let $MIX^e_{G_d,s}(t)$ be the share of source $s \in S$ in the mix prediction of ${G_d}$. 
Thus, the energy intensity factors \rew{can be defined} as:
\begin{itemize}
    \item $CI^e_{G_d}(t) = \sum_{s \in S} MIX^e_{G_d,s}(t) \times 
    CI_s $
    \item $EWIF^e_{G_d}(t) = \sum_{s \in S} MIX^e_{G_d,s}(t) \times EWIF_s $
    \item $ELIF^e_{G_d}(t) = \sum_{s \in S} MIX^e_{G_d,s}(t) \times ELIF_s $
\end{itemize}
where $CI_s$, $EWIF_s$ and $ELIF_s$ are specified in table~\ref{tab:intensities}. 

\review{Note that this formulation captures only operational impacts; embodied impacts from the construction, manufacturing, and eventual decommissioning of data center infrastructure are outside the scope of this model (see Section~\ref{sec:background}).}
    
Moreover, let us define ${\cal{V}}$ as the set of virtual machine instances, for each $v \in {\cal{V}}$ it is known the power draw in relation to a certain job $j \in {\cal{J}}$, $P_{v}$ defined 
according to the Cloud Carbon Footprint methodology\footnote{\url{https://www.cloudcarbonfootprint.org/docs/methodology}}.

Finally, let us also define the set of job requests $J$. For each request $j \in J$ it is known: the arrival/origin region/location $o_j \in D$,
the regions in which data is available ${\cal{D}}[j] \subseteq {\cal{D}}$, data size 
$s_{j} \in \mathbb{N}$, number of nodes $n_{j} \in \mathbb{N}$, the expected utilization $u_{j} \in \mathbb{N}$, the VM instance $v_{j} \in {\cal{V}}$, the expected runtime $r_{j} \in \mathbb{Z}$, the arrival time $a_{j} \in {\cal{T}}$, and the deadline $t_{j} \in {\cal{T}}$.

\rew{The aim is} to optimize footprints associated with the job requests scheduling, considering both data migration and execution, by leveraging spatio-temporal shifting approaches.

\begin{figure*}
    \centering
    \includegraphics[width=\linewidth]{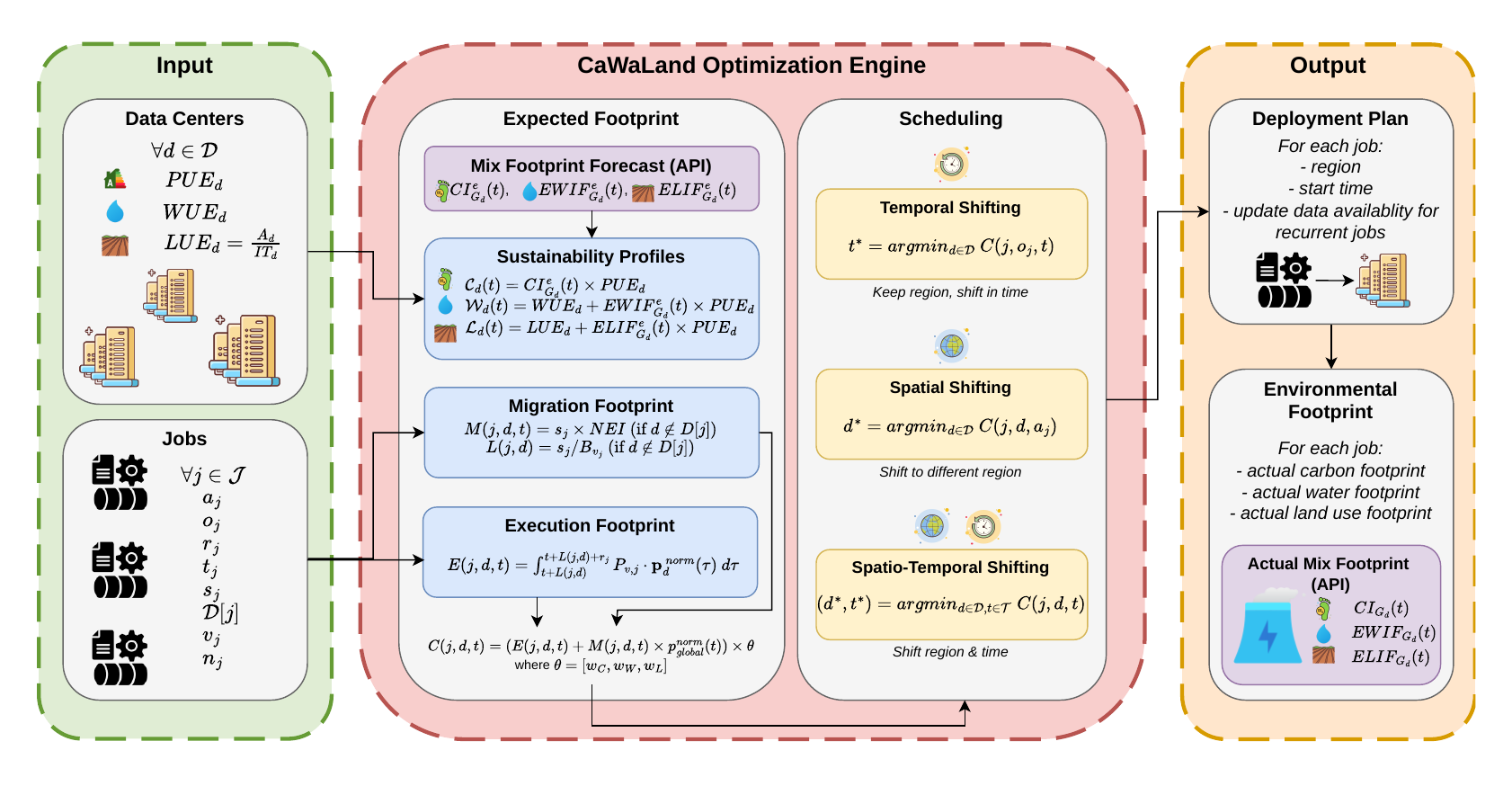}
    \caption{\review{Architecture}}
    \label{fig:architecture}
\end{figure*}

\review{Figure~\ref{fig:architecture} depicts the architecture of the proposed scheduler. 
The diagram shows how job requests (arrival time, origin region, data size and availability, VM configuration, runtime, and deadline) enter the system alongside data center efficiency parameters. The centralized scheduling engine processes these inputs through the cost function calculation, which applies normalization and weighting of the three environmental factors, before generating the final scheduling decisions, applying one of the shifting strategies (spatial, temporal, or spatio-temporal).}

\subsection{Evaluating Environmental Footprints}
\label{sec:footprints}
Firstly, a methodology \rew{is needed} to evaluate the footprints of executing a request in a certain region at a certain time. 
\rew{To do so, it is necessary to predict} the associated footprint of computing a job request $j$ in region/data center $d$ at time $t$. 
If the data related to the request is not available in $d$, data migration is required, and it is necessary to account for its environmental impact.

To account for migration footprint overhead, data transmission energy is estimated as follows:
\begin{equation}
    M(j, d, t) = 
    \begin{cases}
        s_j \times NEI & \text{if } d \notin {\cal{D}}[j] \\
        0 & \text{otherwise}
    \end{cases}
\end{equation}
where  $NEI$ kWh/GB is the Network Energy Intensity, as proposed in~\cite{10.1111/jiec.12630}.

To calculate the execution footprint, executions start \rew{times are needed}.
\rew{Since the scheduling execution time is negligible, it is assumed that if data is not migrated, the execution start would be at the arrival time.} If the data needs to be migrated, data migration latency \rew{is computed} as follows:
\begin{equation}
   L(j,d) =
   \begin{cases} 
         \frac{s_j}{B_{v_j}} & \text{if } d \notin {\cal{D}}[j] \\
        0 & \text{otherwise}
    \end{cases}
\end{equation}

The execution footprint is defined as follows:
\begin{equation}
    E(j,d,t) = \int_{t + L(j,d)}^{t + L(j,d) + r_j} 
        P_{v,j} \cdot \mathbf{p}^{\,norm}_d(\tau)\  d\tau
\end{equation}
where $\mathbf{p}^{\,norm}_d(t) \in \mathbb{R}^3$ is the normalized profile vector of region $d$ 
for carbon, water, and land use.
\rew{Furthermore,} the normalization operator applied component-wise \rew{is defined as follows:}
\[
\mathcal{N}(\mathbf{p}_d(t)) = \frac{\mathbf{x} - \min_{d, t} \mathbf{p}_d(t)}{\max_{d,t} \mathbf{p}_d(t) - \min_{d,t} \mathbf{p}_d(t)} 
\]
Thus, the regional normalized profile is
\(
\mathbf{p}^{\,norm}_d(t) = \mathcal{N}(\mathbf{p}_d(t))
\)

\rew{To account for migration footprint, global intensities are considered and defined} as the normalized (over time) average intensities, namely:
\(
\mathbf{p}^{\,norm}_{global}(t) = 
\mathcal{N}\!\left( \, \mathrm{avg}_{d} \, \mathbf{p}_d(t) \, \right).
\)
Finally, the overall environmental cost of scheduling the job \rew{is computed as follows:}.
\begin{equation}
C(j,d,t) =  (E(j,d,t)  +  M(j,d,t) \times \mathbf{p}^{norm}_{global}(t) ) \times \mathbf{\theta}
\end{equation}
where $\theta$ is the weight vector to linearly combine the three impact factors, namely:
\(
    \mathbf{\theta} =
        \begin{pmatrix}
        w_{\cal{C}} &
        w_{\cal{W}} &
        w_{\cal{L}} \\
        \end{pmatrix}
\).

\subsection{Scheduling Methods}\label{sec:scheduling}
The following shifting methods \rew{are considered}:
\begin{itemize}
    \item \emph{Spatial Shifting}: jobs should be executed where their overall impact is minimized (considering data migration when required).
    \begin{equation}\label{eq:regional}
        d^*, t^* = argmin_{d_\in {\cal{D}}} C(j, d, a_j),\ a_j
    \end{equation}
    \item \emph{Temporal Shifting}: jobs should be executed when their overall impact is minimized.
    \begin{equation}\label{eq:temporal}
         d^*, t^* = o_j,\  argmin_{t\in {[a_j,t_j-r_j]}} C(j, o_j, t)
    \end{equation}
    \item \emph{Spatio-Temporal Shifting}: optimize both when and where the jobs' overall impact is minimized.
    \begin{equation}\label{eq:combined}
         d^*, t^* = argmin_{d \in {\cal{D}},\  t\in {[a_j,t_j-(r_j+L(j,d))]}} C(j, d, t)
    \end{equation}
\end{itemize}
For methods including spatial shifting, \rew{it is assumed} that once data is transferred to another region, the origin location persistently stores the data. This \rew{assumption is} for periodic jobs, data will be available in multiple regions without transferring it multiple times.
As a baseline, a method in which jobs are executed where the request originates \rew{is considered}. Namely, the \emph{Local} baseline:
\begin{equation}\label{eq:local}
    d^*, t^* = o_j,\ a_j
\end{equation}

\subsection{Experiment Design}\label{sec:experiment_design}
In this work, \rew{the focus is} on two types of workloads. \rew{Namely}: a user-facing, latency-sensitive workload (FaaS) and a delay-tolerant batch processing workload (big data). For each workload type, real traces \rew{are considered} and \rew{a real-world scenario is derived} to simulate their scheduling and execution.


\paragraph{Aim}
The aim of our experiments is to investigate environmentally conscious cloud workload shifting through realistic simulations, not to compare real-world providers. Our results may not accurately reflect the impact of real-world workloads on AWS or Azure data centers, because, although our scenarios are inspired by published data where possible, assumptions and approximations are made where exact data was not available.
\review{
This simulation-based methodology was chosen because it allows us to explore a wide range of scheduling strategies, weight configurations, seasons and prediction-error levels across large, geographically distributed infrastructures. It would be impractical and environmentally counterproductive to replicate this scale and breadth through the real-world deployment of every configuration studied.
}

\review{
\paragraph{State-of-the-art Comparison}
Our carbon-only temporal-shifting configuration follows the same underlying  principle as \emph{Let's Wait Awhile}~\cite{10.1145/3464298.3493399}: both defer job execution within a bounded delay-tolerance window to align with periods of lower carbon intensity, thus solving the same one-dimensional temporal optimization problem. While the two studies use different delay-tolerance values, but this does not affect the comparability of the underlying mechanism. This strategy will be referred to as "T (dt*) - carbon-only (WaitAwhile)".

Similar to our carbon-only spatial-shifting strategy, GreenCourier~\cite{10.1145/3631295.3631396} allocates workloads to the region with the lowest carbon impact. However, GreenCourier makes decisions based on real-time marginal carbon intensity, whereas forecasted average carbon intensity is used in this study. Furthermore, GreenCourier does not consider for the environmental cost of migrating job data, or the energy required to run a job, nor how this energy translates into a carbon footprint over time. To enable a fair comparison, a heuristic inspired by GreenCourier that decides where to execute a job based on the current average carbon intensity was implemented. This strategy will be referred as "S - carbon-only ($\sim$ GreenCourier)"

WaterWise~\cite{jiang2025waterwise} 
jointly minimizes the carbon and water footprints of geographically distributed job scheduling. Their experimental setup is comparable to our equally weighted carbon-and-water configuration ($w_C=0.5, w_W=0.5, w_L=0$). However, there are several differences: WaterWise normalizes each footprint by the maximum value achievable for a given job. In contrast, in this work the normalization is computed across the full observed range of regions and time steps. Additionally, WaterWise incorporates a Water Scarcity Factor that scales the  water footprint by regional water stress, as well as a history-learner term that accounts for recent footprint trends. Neither of these factors are modeled in our formulation. Finally, WaterWise solves its objective via Mixed Integer Linear Programming with explicit capacity and delay-tolerance constraints. In contrast, a direct $\arg\min$ search over discretized regions is performed in this work. 
These differences mean that our results should not be interpreted as a reproduction of WaterWise's reported savings. The shared normalized-weighted-sum structure and matching default weights justify the correspondence that can be drawn between the two configurations. For this reason, it is suggested that the strategy corresponding to the equally weighted carbon-and-water configuration is inspired by WaterWise and this strategy will be referred as "S - carbon-water ($\sim$ WaterWise)".

}

\paragraph{Simulation Overview}

The discrete event simulation detailed in Algorithm~\ref{alg:iterative_scheduling} \rew{has been implemented}. At each time step, we update the sustainability profiles of data centers and new job requests ready to be scheduled. 
Then each job can be scheduled according to the shifting approach previously defined. 
Finally, since \rew{it is assumed} that once data is transmitted to a region, it is persistently stored there, the data availability set of the job is updated for future scheduling.

{
\begin{algorithm}
\caption{Simulation Procedure}
\begin{algorithmic}[1]\label{alg:iterative_scheduling}
\STATE \emph{Input:} ${\cal{D}}$ data centers, 
\(\texttt{scheduling\_method()} \in \{eq. \eqref{eq:regional},eq.\eqref{eq:temporal}, eq.\eqref{eq:combined}, eq.\eqref{eq:local}\}\)

\FOR{$t \in {\cal{T}}$}
    \STATE Get sustainability profiles $p_d(t)$ for each $d \in {\cal{D}}$
    \STATE Get jobs ready to be scheduled at time $t$, $J(t)$.
    \FOR{$j$ in $J(t)$  \emph{parallel}}
        \STATE $d^*, t^* = \texttt{scheduling\_method()}$  
        \IF{\(\texttt{scheduling\_method()} \in  \{eq.\eqref{eq:regional}, eq.\eqref{eq:combined}\}\)}
        \STATE ${\cal{D}}[j] = {\cal{D}}[j] \cup \{d^*\}$
        \ENDIF
    \ENDFOR
\ENDFOR
\STATE \emph{Output:} Final schedule over horizon $H$
\end{algorithmic}
\end{algorithm}
}

\paragraph{Scenario 1 - Azure and FaaS} 
The first scenario is based on Azure regions and a FaaS workload. 
The regions selected for this scenario are "swedencentral", "southcentralus", "centralus", and "eastus2". 
Table~\ref{tab:azure_regions} reports regions' characteristics. 
PUE and WUE values were taken from Azure issued sustainability reports.
The land occupation data stems from online reports.
As for the workload, the 2019 Azure Functions dataset \rew{was chosen}. For the data sources see~\ref{online_res}.
The source data is composed of two sets of daily CSV files: one detailing per-minute invocation counts for each function, and another providing daily summary statistics of function execution times, including averages and key percentiles (0th, 1st, 25th, 50th, 75th, 99th, and 100th). The first seven days of this dataset \rew{was utilized} to synthesise 100,000 requests of function execution a day. The selection of a function for each request was probabilistic and weighted by the function's total daily invocation count, such that more frequently invoked functions appeared more often in the trace, mirroring real-world usage patterns. 

Upon selecting a function, a synthetic runtime \rew{was sampled} by performing linear interpolation on its known daily execution time percentiles. The arrival time for the request, resolved to the minute-of-day, was similarly sampled from the function's empirical per-minute invocation distribution. Finally, each request was assigned a data center location from a predefined set. 
A single virtual machine configuration (2 vCPU and 4 GB of RAM) is considered for this scenario, as specified in the paper~\cite{azurefaas}.
           
For this scenario, only the \emph{Local} baseline (L) and the \emph{Spatial Shifting} (SP) approach \rew{are compared}.
That is, \rew{it is assumed} that the functions are tolerant to the latency required to change regions but not to be delayed in time. For this reason, \emph{Temporal Shifting} (T) and \emph{Spatio-temporal Shifting} (STP) scheduling methods \rew{are excluded}.
Thus, there are no deadlines for this scenario.
{\scriptsize
\begin{table}[h!]
\centering
\begin{tabular}{|l|l|p{1cm}|p{1cm}|p{1cm}|}
\hline
\emph{Region} & \emph{Location} & \emph{PUE} & \emph{WUE} & \emph{Land Occup. ($m^2$)} \\
\hline
swedencentral & Sweden & 1.172 & 0.16 & 1300000 \\
southcentralus & Texas & 1.307 & 1.82 & 43664\\
centralus & Iowa & 1.16 & 0.19 & 37904\\
eastus2 & Virginia & 1.144 & 0.17 & 102193\\
\hline
\end{tabular}
\caption{Azure data by Region for Scenario 1.}\label{tab:azure_regions}
\end{table}
}
\paragraph{Scenario 2 - AWS and Big Data}

The second scenario is based on AWS and Big Data workloads.
The regions selected for this scenario are: "us-east-1", "us-west-1", "eu-central-1", "eu-west-2", and "eu-north-1".
Table~\ref{tab:aws_regions} reports regions' characteristics. PUE and WUE values were taken from AWS website (see~\ref{online_res})

The land occupation data is estimated based on the 10-k form released by AWS in 2024 (see~\ref{online_res}),
in which it is stated that 
more than 450,000 $m^2$ were owned/rented by AWS for data centers. 
The estimation is also based on declared availability zones for each region. Namely, the estimated land occupation for a given region is calculated by taking the number of availability zones (AZs) in that region, dividing it by the total number of AZs (117), and then multiplying by the total land occupation of all data centers.

As for the workload, requests from available experimental Spark traces (see~\ref{online_res}) \rew{were synthetically generated}.
50k requests over a week \rew{were randomly generated}. For the arrival pattern, 50\% of requests \rew{are generated} as periodic (in line with literature analysis: about 43\%~\cite{8276798}, about 60\%~\cite{10.5555/3026877.3026887}) and the rest as ad-hoc jobs. 
The arrival of ad-hoc jobs follows a Poisson distribution with \(\lambda = target_{np} / n_{days} * minutes per day \approx 2.5\). The periodic jobs' first instance arrival time is randomly picked within the first 12 hours, and their periodicity is set uniformly at random among \(\{2, 4, 8, 12\}\) hours.
For each request, the origin location is picked uniformly at random from the regions considered for this scenario. The number of nodes, the VM instance, and the expected runtime are the ones reported in the traces. Table~\ref{tab:aws_vm} summarizes the VM instances considered in our experiments, and their specifications are set accordingly (memory, cpu, 
and bandwidth 
, see~\ref{online_res}). 

\review{
Note that the synthetically generated aspects of the trace (arrival timing and origin location) do not alter the core, empirically-derived workload characteristics: the number of nodes, VM 
instance type, and expected runtime for each job are drawn directly from the original Spark execution traces~\cite{scoutpapaer}. Arrival timing and origin location were necessarily synthesized because, to the best of our knowledge, no publicly available workload trace records the execution of a large-scale big data workload across multiple, geographically distributed cloud sites.
}

For this scenario, all of the scheduling methods defined above \rew{are considered. A} variation of \emph{Spatial Shifting} \rew{is also introduced}, in which \rew{data transferred when migrating is not persistently stored} (S). 
For \emph{Temporal Shifting} (T) and \emph{Spatio-Temporal Shifting} (SPT), the deadline is either its periodicity or, for ad hoc jobs, a tunable delay tolerance parameter.
{\scriptsize
\begin{table}[h!]
\centering

\begin{tabular}{|l|l|l|l|p{1cm}|p{1cm}|p{1cm}|}
\hline
\emph{Region} & \emph{Location} & \emph{PUE} & \emph{WUE} & \emph{Land Occup. ($m^2$)} \\
\hline
us-east-1 & Virginia &1.15 &0.12 & 233101 \\
us-west-1 & California & 1.17 & 0.51 & 116550 \\
eu-central-1 & Frankfurt & 1.35 & 0.01 & 116550  \\
eu-west-2 & London & 1.11 & 0.04 & 116550 \\
eu-north-1 & Stockholm & 1.10 & 0.02 & 116550 \\
\hline
\end{tabular}

\caption{AWS data by Region for Scenario 2.  }\label{tab:aws_regions}
\end{table}
}
{\scriptsize
\begin{table}[]
    \centering

    \begin{tabular}{|ll|c|c|p{2.2cm}|}
        \hline
        \multicolumn{2}{|c|}{Type}  & vCPU & Mem (GB) & Bandwidth (Mbps)\\
        \hline
        c. &l, xl, xxl &  2, 4, 8 & 4, 8, 16 & 500, 750, 1000 \\
        \hline
        m4. & l, xl, xxl &  2, 4, 8 & 8, 18, 34 & 450, 750, 1000 \\
        \hline
        r4. & l, xl, xxl & 2, 4, 8 & 16, 32, 65 & 450, 850, 1700 \\
        \hline
    \end{tabular}

\caption{AWS VM instance types and specifications}\label{tab:aws_vm}
\end{table}
}


\subsection{Experimental Setup}
\label{sec:setup}

\paragraph{Hardware and Software Environment}
The experiments were conducted on the following system:
\begin{itemize}
    \item \emph{CPU:} Intel(R) Core(TM) i9-10920X CPU @ 3.50GHz 
    \item \emph{RAM:} 260 GB 
    \item \emph{Operating System:} Ubuntu 24.04.3 LTS 
\end{itemize}

\paragraph{Grid Data and Intensities} 
Since multiple regions \rew{are considered}, publicly available data from different grid operators (see~\ref{online_res}) \rew{was collected}.
The historical mix data is used to compute the actual environmental impact of the workload execution.
Meanwhile, noised mix data is used to simulate the prediction of the grids' mix that the schedulers optimize their decisions on.
The data sources are not standardized, thus \rew{data was harmonized} to a common format \rew{considering} a granularity of 1 hour. Thus, finer-grained time series were reshaped considering the mean value for each hour.
Table~\ref{tab:data_sources} depicts the relationship between grids and cloud providers' regions.

\begin{table*}[ht]
\centering

\begin{tabular}{|c|cccc|}
\hline
\multirow{3}{*}{EU} & sw & uk & de & \\
\cline{2-5}
& \makecell{ eu-north-1\\ swedencentral }
        & eu-west-2 
        & eu-central-1 
        &\\
\hline
\multirow{3}{*}{USA} & ercot & miso & pjm & caiso \\
\cline{2-5}
 & southcentralus 
        & centralus 
        & \makecell{ us-east-1\\ eastus2} 
        & us-west-1 \\
\hline
\end{tabular}

\caption{Data sources for grid mixes}\label{tab:data_sources}
\end{table*}

To incorporate uncertainties into the grid mix data, synthetic noise \rew{is introduced}. 
Specifically, for each time step and region, the renewable share is perturbed by an error term $\epsilon_r$ sampled from a normal distribution $\mathcal{N}(0, \sigma_r)$, where $\sigma_r = mae \cdot \frac{1}{\sqrt{2/\pi}}$ and $mae$ is the target mean absolute error. 
The error $\epsilon_r$ is then distributed among the renewable sources (solar, wind, geothermal, hydro, biomass) according to a probability distribution designed to reflect their prediction difficulty: solar gets 45\% of error, since it is the hardest source to estimate~\cite{alfadda2017hour}; wind gets 30\%~\cite{okumus2016current}; both hydro~\cite{zayas2025new} and geothermal~\cite{bilgili2022one} get 10\%; and biomass get 5\% of the error, being the most controllable among the renewable sources. 
Each renewable source $i$ receives a noise $\epsilon_i = \epsilon_r \cdot w_i$, where $w_i$ is the weight for source $i$.
This approach is an attempt to simulate realistic prediction errors and reflects the varying uncertainty associated with each energy source.
For non-renewable sources, a uniform noise is applied to each share to maintain the total mix balance. 

Finally, in our simulations, to compute the average intensities of the mixes, we consider the intensity coefficients reported in  Table~\ref{tab:intensities}.
Carbon intensity values correspond to the median of the lifecycle emissions reported in IPCC Annex III, Table  IPCC A.III.2 table (page 1335)~\cite{ipcc2014_ar5_annexIII}. 
For the oil carbon intensity, which is missing in this source, another coefficient\cite{owid_safest_sources} \rew{is considered}.
Water intensity values were obtained from the Water Consumption Factors presented in Tables 1 and 2 (pages 12-13) of the NREL technical report on energy-related water consumption~\cite{nrel2011_water}. 
Land use intensity values were extracted from a comprehensive survey~\cite{plos2022_landuse}. The survey does not provide a value for oil, but the authors assume it to be comparable to gas, so \rew{the same holds for this work.}
For all the factors, when multiple options were presented for the same energy source, the average of median intensity values \rew{was taken}.
Note that \rew{it was not possible to} find any source for the oil EWIF; thus, \rew{it is assumed} the mean of the other sources' intensity values. As well as for the unknown share.

\begin{table}[h!]
\centering

\begin{tabular}{p{1.5cm}ccc}
\hline
\multirow{2}{1.5cm}{\emph{Energy Source}} 
& \emph{CI} & \emph{EWIF} & \emph{ELIF} \\
 & \textit{$gCO2e/kWh$} & \textit{$l/kWh$} & \textit{$ha/TWh$}  \\
\hline
Nuclear    & 12     & 1.957   & 7.1 \\
Geothermal & 38     & 5.827   & 45 \\
Biomass    & 485    & 1.145   & 29065  \\
Coal       & 820    & 1.8   & 1000 \\
Wind       & 11.5   & 0  & 6065  \\
Solar      & 38.67  & 1.385   & 1650  \\
Hydro      & 24     & 17 & 650 \\
Gas        & 490    & 1.099   & 1155 \\
Oil        & 720    & n/a (3.776) & 1155 \\
Unknown    &  n/a (293.24)    & n/a (3.776) &  n/a (4532) \\
\hline
\end{tabular}

\caption{Impact coefficients for various electricity generation technologies.
\emph{Sources:}  
Carbon: IPCC~\cite{ipcc2014_ar5_annexIII}, Our World in Data~\cite{owid_safest_sources} (oil).  
Water: NREL~\cite{nrel2011_water}.  
Land use~\cite{plos2022_landuse}.
}\label{tab:intensities}
\end{table}

\paragraph{Other Parameters}  

For both scenarios the mid-season one-week horizon \rew{was considered}, from the midnight of the 15th to the midnight of the 22nd, considering the UTC time zone, and varied the week in different seasons, namely: winter (January), 
spring (April), 
summer (July), 
autumn (October).
And the step $\delta_t$ is set to one minute.

To account for the uncertainty in the prediction of the workload, different values of MAE \rew{were considered}, namely: 0.05, 0.1, 0.15, and 0.2.
For statistical significance, each simulation is repeated 6 times with different random seeds, namely: 0, 1, 2, 3, 4, 5.

The factors' weights are set to: [1.0, 0.0, 0.0], only carbon footprint is optimized, [0.0, 1.0, 0.0], only water footprint is optimized, [0.0, 0.0, 1.0], only land footprint is optimized, [0.333, 0.333, 0.334], all footprints are optimized equally.

Moreover, to account for the energy consumption of the network, \rew{the network energy intensity was set to} 0.06 kWh/GB~\cite{nei}.
Finally, \({P^{IT}_d}\) \rew{was set} to the estimate provided by EIA. Assuming the IT power draw to be $100$ MW, IT power consumption \rew{is assumed to be} $8760 \times 100 \times 10^3$  kWh over one year.

\section{Results}\label{sec:results}
In this section, the results of our experiments \rew{are presented}.
\subsection{Scenario 1: Azure and FaaS}
\begin{figure*}
    \centering
    \includegraphics[width=\linewidth]{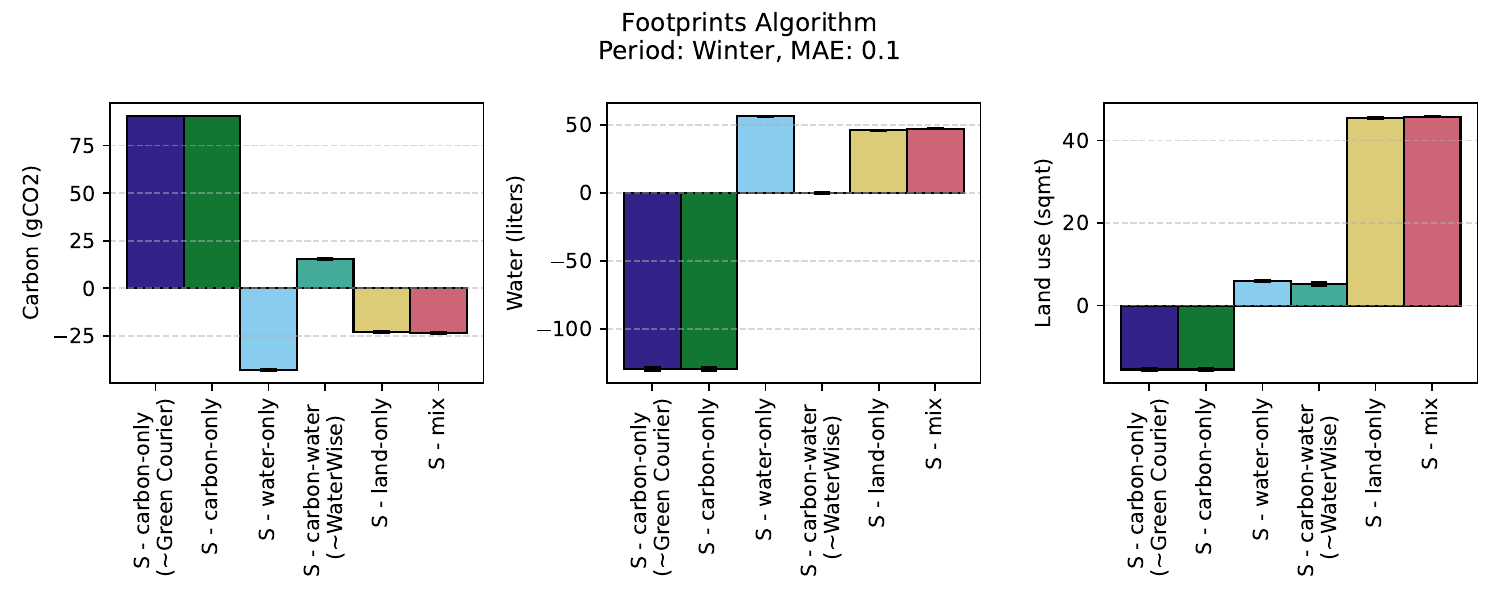}
    \caption{Spatial shifting for Scenario 1; showing a $45-85\%$ improvement over local baseline during a winter week, with an MAE of $10\%$.}
    \label{fig:azure_R_winter_0.1_improvement}
\end{figure*}

For this first scenario, the results of the spatial shifting approach optimizing for all factor weight combinations \rew{are provided}.
\paragraph{Spatial Shifting}
Figure~\ref{fig:azure_R_winter_0.1_improvement} shows the footprint improvement with respect to the local baseline of spatial shifting. This experiment considered the winter week and an error of 10\% on the renewable share prediction. All factor weights' combinations defined above \rew{are compared}.

The results demonstrate that regional shifting of FaaS workloads across Azure regions leads to noticeable improvements in carbon efficiency. The plots show that when optimizing for a particular metric (carbon, water, or land use footprint), a $45-85\%$ improvement in that metric \rew{can be achieved}. When considering the metric for which the optimization was performed, the carbon footprint benefits the most, with up to an $85\%$ reduction in emissions compared to the local baseline. This significant reduction reflects the negligible migration cost of FaaS workloads.
\review{
For both \textit{ S - carbon-only} and S - carbon-only ($\sim$ GreenCourier), a}
reduction of about 85\% in carbon footprint leads to an increase of about 40\% in water footprint. The land use footprint also increases, but to a lesser extent, having observed about a 25\% increase. 

\review{\textit{S - carbon-only ($\sim$ GreenCourier)} achieves the same results as the \textit{S - carbon-only} strategy due to the short runtime nature of FaaS.}

Similarly, optimizing for water footprint (about 50\% reduction) leads to a significant increase in carbon footprint (about 125\% increase), but the land use footprint decreases proportionally to the water footprint reduction (about 50\% reduction). On the other hand optimizing for land use footprint (about 45\% reduction) leads to a moderate increase in carbon footprint (about 15\% increase) and a small reduction in water footprint (less than 10\% reduction).

\begin{figure*}
    \centering
    \includegraphics[width=\linewidth]{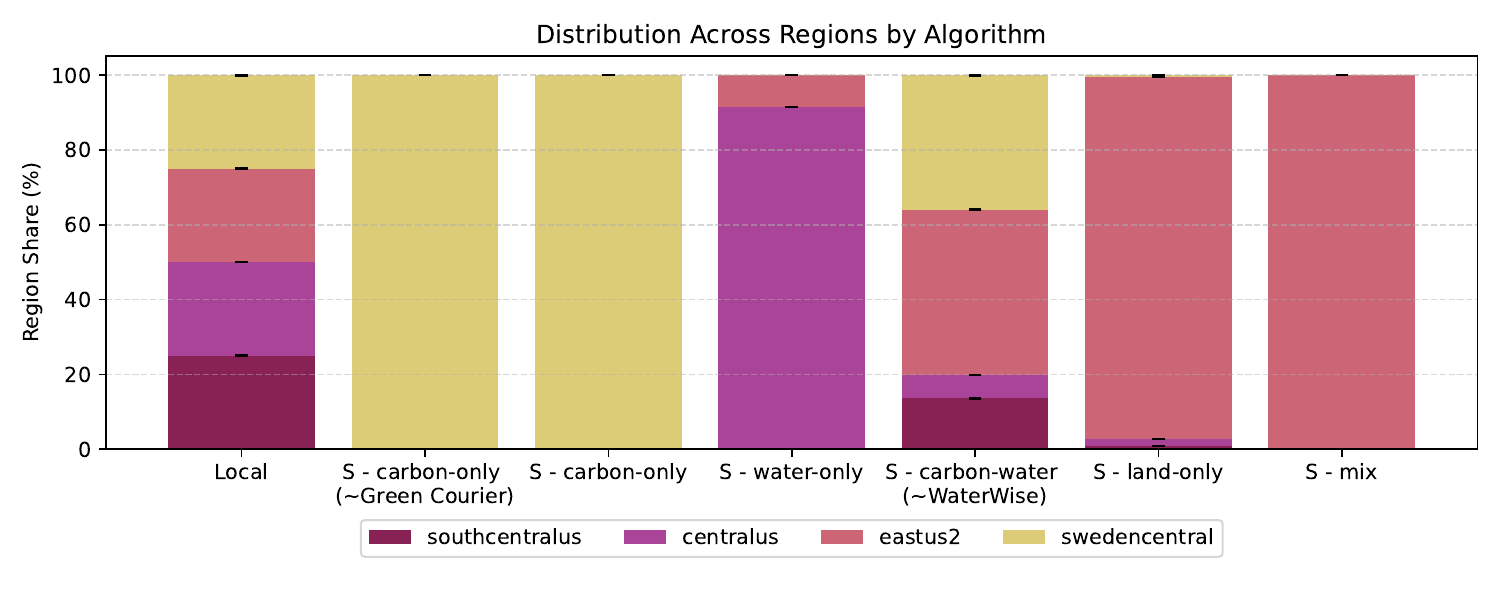}
    \caption{Request distribution across different regions for different criteria for Scenario 1 during a Winter week, with an MAE of $10\%$.}
    \label{fig:azure_load}
\end{figure*}

Figure~\ref{fig:azure_load} shows the request distribution across the regions.

\begin{figure*}
    \centering
    \includegraphics[width=\linewidth]{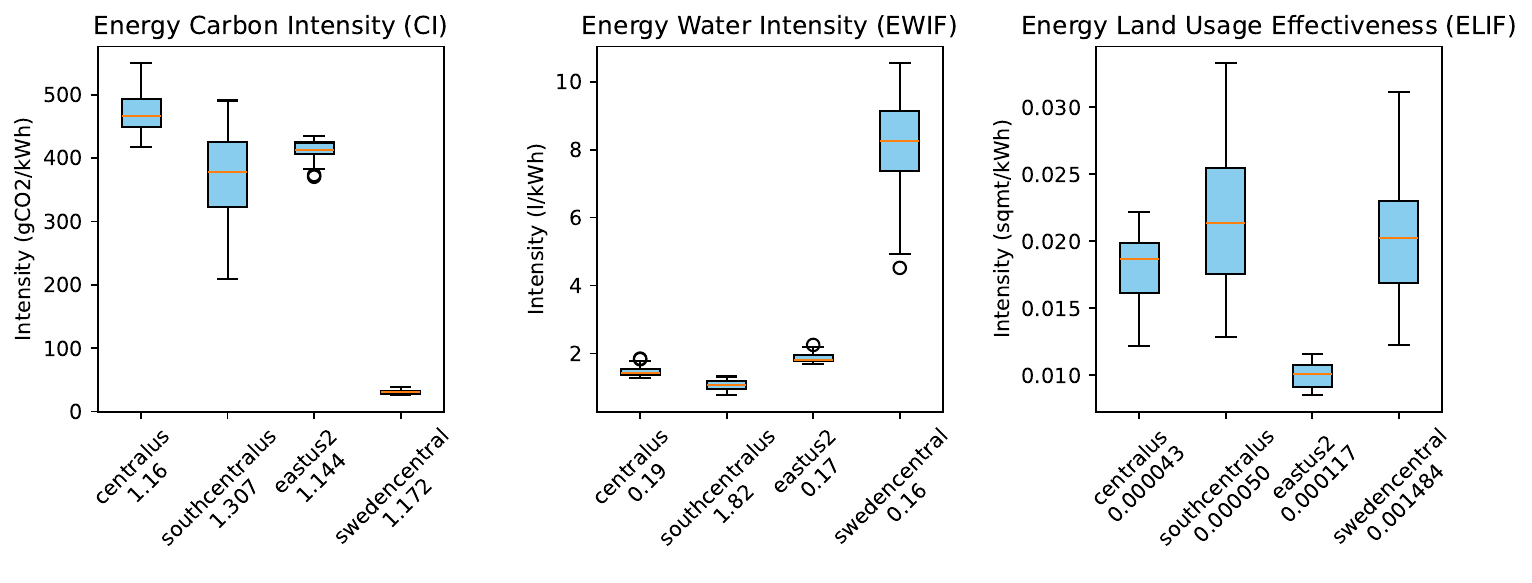}
    \caption{Grid intensities for each factor, and each region for Scenario 1 over a winter week, with an MAE of $10\%$.}
    \label{fig:azure_grid_intensities}
\end{figure*}

To understand this distribution, consider Figure ~\ref{fig:azure_grid_intensities}, which simultaneously reports intensity values over the winter week and the efficiency metrics for each factor for each region.

When optimizing for carbon, all the requests are scheduled to "centralus" which is not the most energy-efficient region (PUE 1.172), and it has the grid with the lowest carbon intensity. 
When optimizing for water, a small percentage of requests go to "eastus2" and most of the requests are scheduled to "centralus", which is not the most water-efficient region (WUE 0.19) nor is it the one with the smallest grid water intensity. However, it is the region with the lowest impact on water when considering both data center efficiency and grid intensity. 
Finally, when optimizing for land use, nearly all the requests are scheduled to "eastus2", which, even if it is not the region with the least land occupation, its grid has a lower land use intensity.

\begin{figure*}
    \centering
    \includegraphics[width=\linewidth]{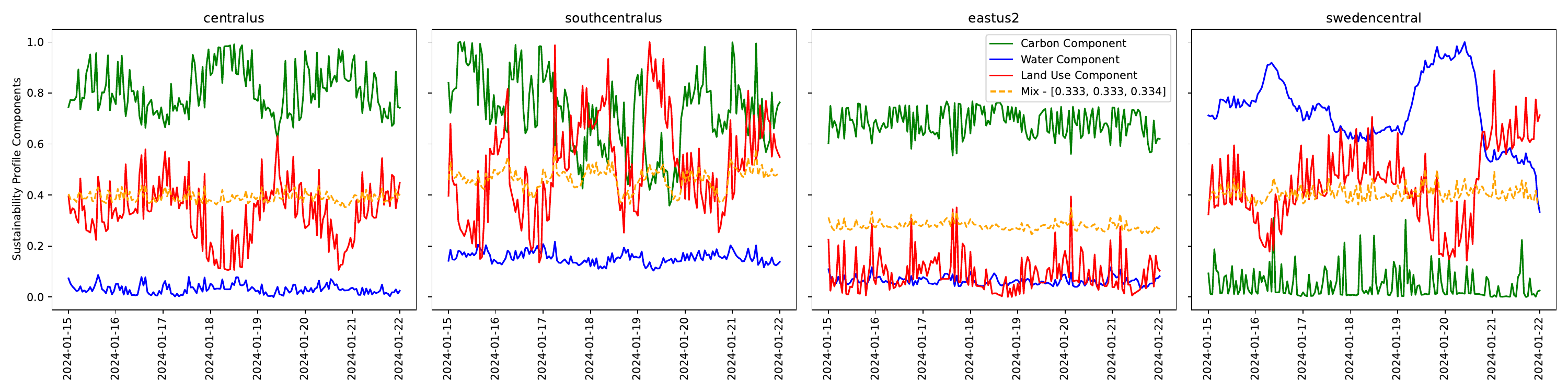}
    \caption{\review{Sustainability Profile components for Scenario 1 over a winter week, with an MAE of $10\%$.}}
    \label{fig:azure_sust_profiles}
\end{figure*}

\review{
Figure~\ref{fig:azure_sust_profiles} shows the sustainability profiles. For each region, all the components are plotted: green for carbon, blue for water and red for land use. The linearly scaled profile by the mix configuration $\theta = (0.333, 0.333, 0.334)$ is plotted in orange. The average distance between the mix configuration signal and the components of the sustainability profile are 0.36, 0.30 and 0.14 respectively for carbon, water and land use. This explains why the mix strategy seems to be particularly influenced by the land use component.
}

\begin{figure*}
    \centering
    \includegraphics[width=\linewidth]{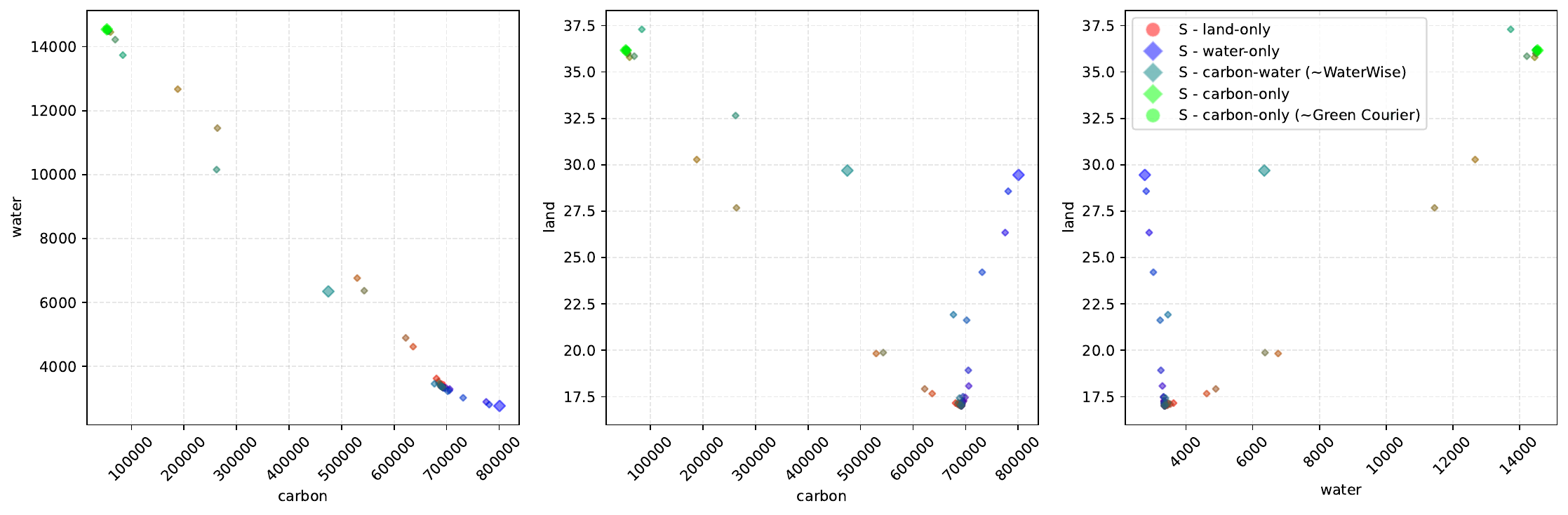}
    \caption{\review{2D - Pareto Analysis for Scenario 1 over a winter week, with an MAE of $10\%$.}}
    \label{fig:azure_pareto}
\end{figure*}

\review{
\paragraph{Pareto Analysis}
To characterize the trade-offs among carbon, water, and land-use footprints beyond the four representative weight configurations discussed above, we performed a Pareto analysis for Scenario 1. The season is fixed to Winter and the grid error to 0.1. The frontier is approximated by evaluating 66 configurations of weight vectors $\theta = (w_C, w_W, w_L)$. Figure~\ref{fig:azure_pareto} shows the resulting points projected onto the three pairwise objective planes: carbon vs. water, carbon vs. land, and water vs. land. 
Note that the color of the data points reflects the configuration of the theta (red to $w_{\cal{L}}$ , green to $w_{\cal{C}}$ and blue to $w_{\cal{W}}$), and that non-dominated data points are represented as diamonds, while dominated data points are represented as dots.
Each single-objective configuration (carbon-only, water-only, land-only) sits at an extreme of the frontier, reflecting the cost of optimizing for one footprint in isolation. The frontier shows a clear three-way trade-off, in which reducing the carbon footprint tends to increase both the water and land-use footprints, and vice versa.
Moreover, it is clear that no single configuration minimizes all three objectives simultaneously. 
The carbon-water configuration ($\sim$WaterWise) belongs to the frontier, and never sits at the extreme of any axis. The carbon-only configuration ($\sim$GreenCourier) is dominated but by such a tiny amount that it is not really visible from the plot.
}

\paragraph{Sensitivity Analysis}

To understand how seasonality affects the effectiveness of the scheduling methods,  a sensitivity analysis \rew{is performed} considering weeks in the four seasons: winter, spring, summer, and autumn. For this analysis, a fixed error for the grid's renewable share prediction \rew{is selected} (mae = 0.1).
On the other hand, to understand how the prediction error affects the effectiveness of the scheduling methods, a sensitivity analysis considering different values of prediction error \rew{is performed} (mae = \{0.05, 0.1, 0.15, 0.2\}). A fixed week in winter \rew{is used} for this analysis.

From Tables~\ref{tab:azure_sa_season} and~\ref{tab:azure_sa_mix}, it is possible to notice that the prediction error leads to very small differences, almost flat. Indeed, the results indicate that the improvement in reduction in all three metrics remains robust even for higher error levels.

On the other hand, it seems that the effectiveness of spatial shifting varies with seasonal changes in grid intensities, but the overall trend of improvement remains consistent. Land use factors, in particular, demonstrate varying impacts across different seasons (37\%-57\%).

\subsection{Scenario 2: AWS and Big Data}\label{sec:res_aws}

For this second scenario, not only \rew{is} the spatial shifting method \rew{investigated}, but also temporal shifting and spatio-temporal shifting.
The same week in winter and the same factor weights' combinations as in Scenario 1 \rew{were considered}. The prediction error (mae) \rew{was set to } 0.1. For temporal shifting, different values of delay tolerance \rew{were considered, namely} (dt) = \{4, 12, 24, 48\} hours.

\begin{figure*}
    \centering
    \includegraphics[width=\linewidth]{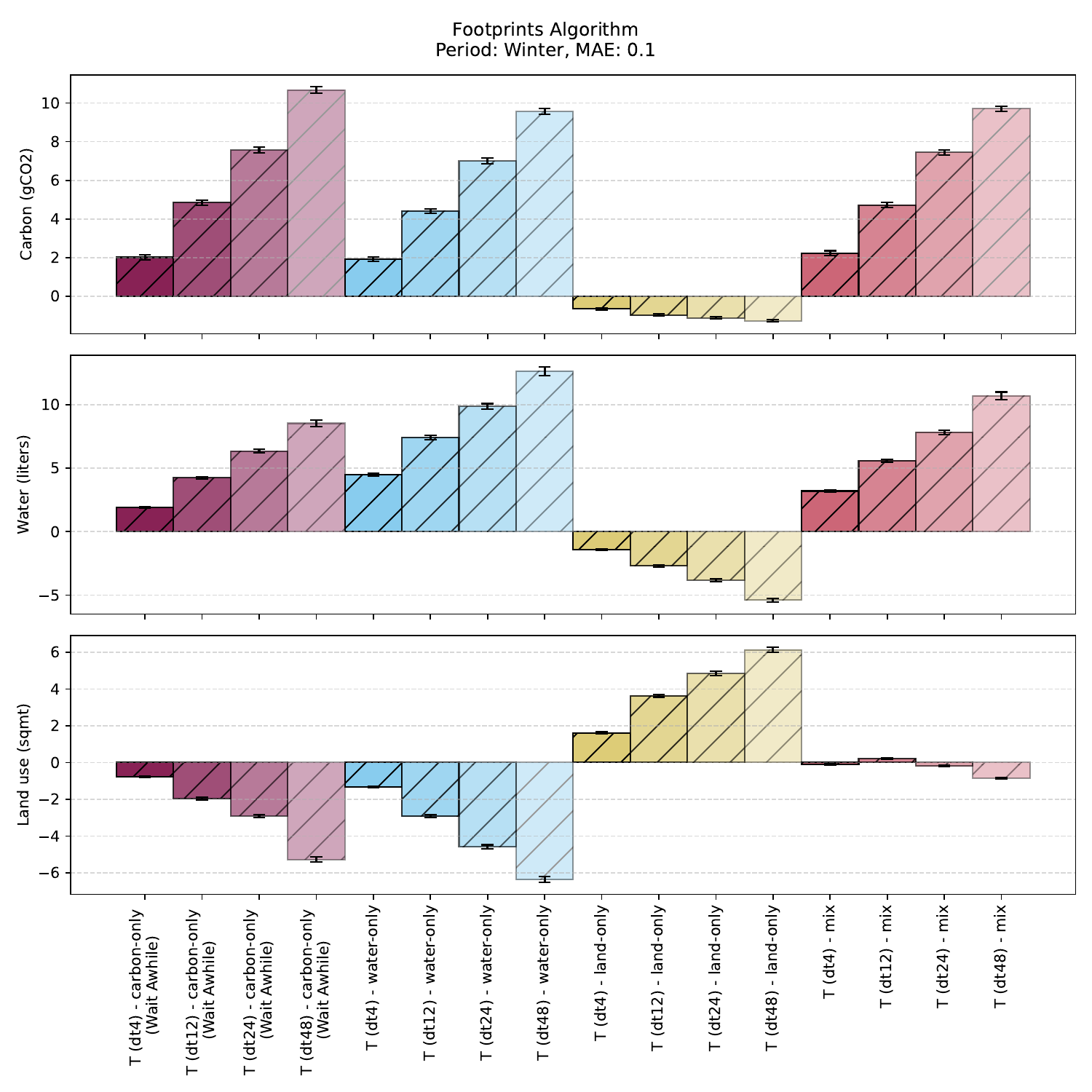}
    \caption{\review{Temporal} shifting for Scenario 2 improvement over local baseline results, with an MAE of $10\%$. Temporal shifting shows $6-12\%$ improvement; spatial shifting comes in at $20-45\%$; while combining both gives the best result of $40-55\%$ improvement over the baseline.}
    \label{fig:aws_winter_0.1_improvement_temporal}
\end{figure*}

\begin{figure*}
    \centering
    \includegraphics[width=\linewidth]{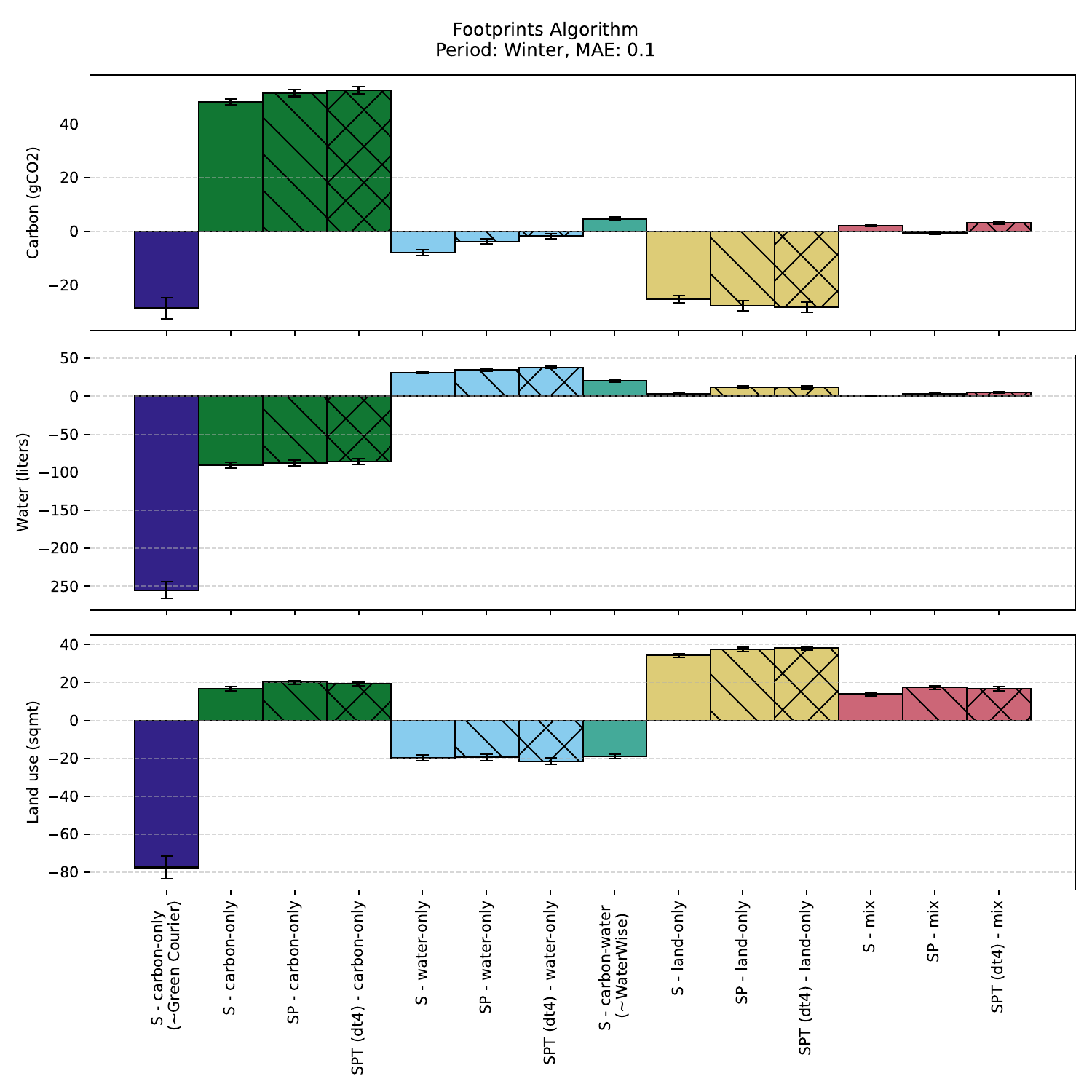}
    \caption{\review{Spatial, and spatio-temporal shifting} for Scenario 2 improvement over local baseline results, with an MAE of $10\%$. Temporal shifting shows $6-12\%$ improvement; spatial shifting comes in at $20-45\%$; while combining both gives the best result of $40-55\%$ improvement over the baseline.}
    \label{fig:aws_winter_0.1_improvement_spatiotemp}
\end{figure*}

\paragraph{Baseline Footprints}
The baseline algorithm (Local) has about 17.500 kilograms of carbon footprint, 200.000 liters of water footprint, and just above 2000 square meters of land use footprint. In these graphs, \rew{is shown} the improvement in percentage of the other algorithms with respect to the baseline.
\review{Figure~\ref{fig:aws_winter_0.1_improvement_temporal} and Figure~\ref{fig:aws_winter_0.1_improvement_spatiotemp} }show the results of our experiments. 
\paragraph{Temporal Shifting}
Firstly, \review{as shown in Figure~\ref{fig:aws_winter_0.1_improvement_temporal},} the temporal shifting approach achieves relatively modest improvements compared to spatial shifting, with reductions ranging from 6\% to 12\% even at the highest delay tolerance of 48 hours.

\paragraph{Spatial Shifting}
The spatial shifting approach (SP) achieves a significant improvement in each metric when optimizing for it \review{(Figure~\ref{fig:aws_winter_0.1_improvement_spatiotemp})}.
In this case, the improvements range from $20-45\%$ footprint reductions over the baseline. 
\begin{figure*}
    \centering
    \includegraphics[width=\linewidth]{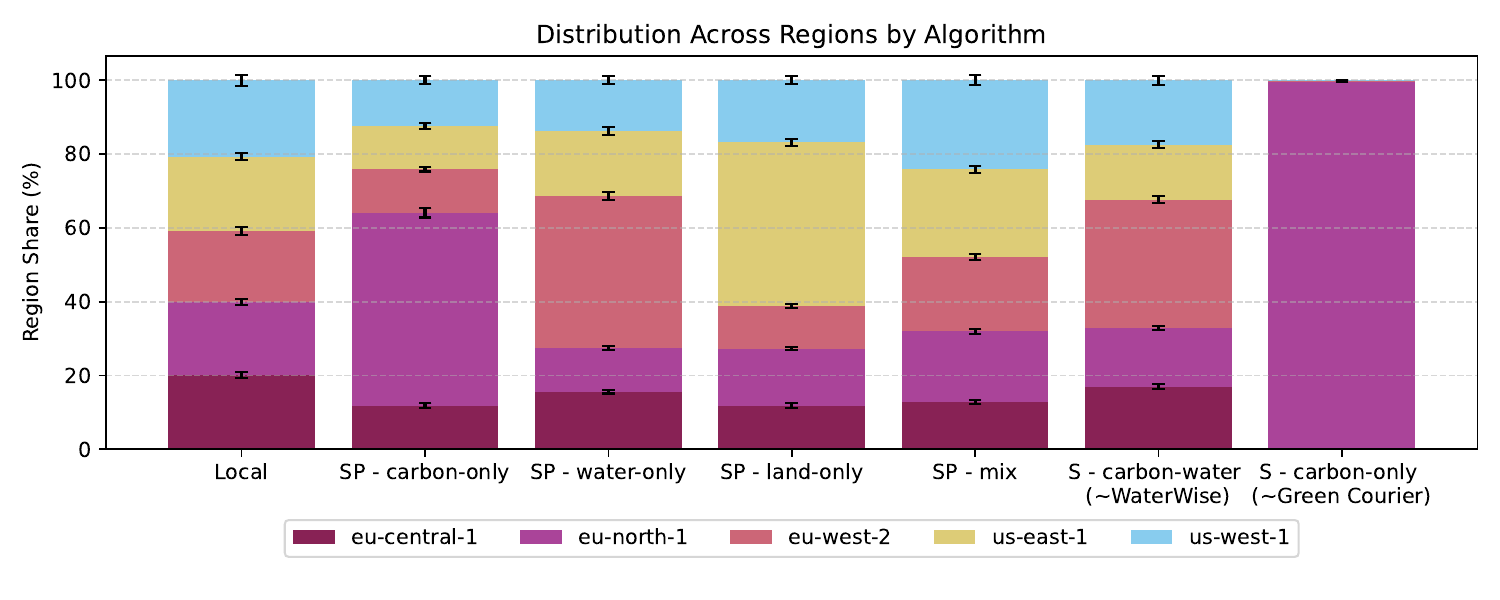}
    \caption{Request distribution across different regions for different criteria for Scenario 2 during a Winter week, with an MAE of $10\%$.}
    \label{fig:aws_load}
\end{figure*}
\review{
Our approach outperforms \textit{S - carbon-only ($\sim$ GreenCourier)}, since ignoring migration footprints and the implications of long-running jobs results in even worse results with respect to the local baseline. \textit{S - carbon-only ($\sim$ Water Wise)} exhibits behavior similar to \textit{S - water-only}, except it does not worsen the carbon footprint.
}

\begin{figure*}
    \centering
    \includegraphics[width=\linewidth]{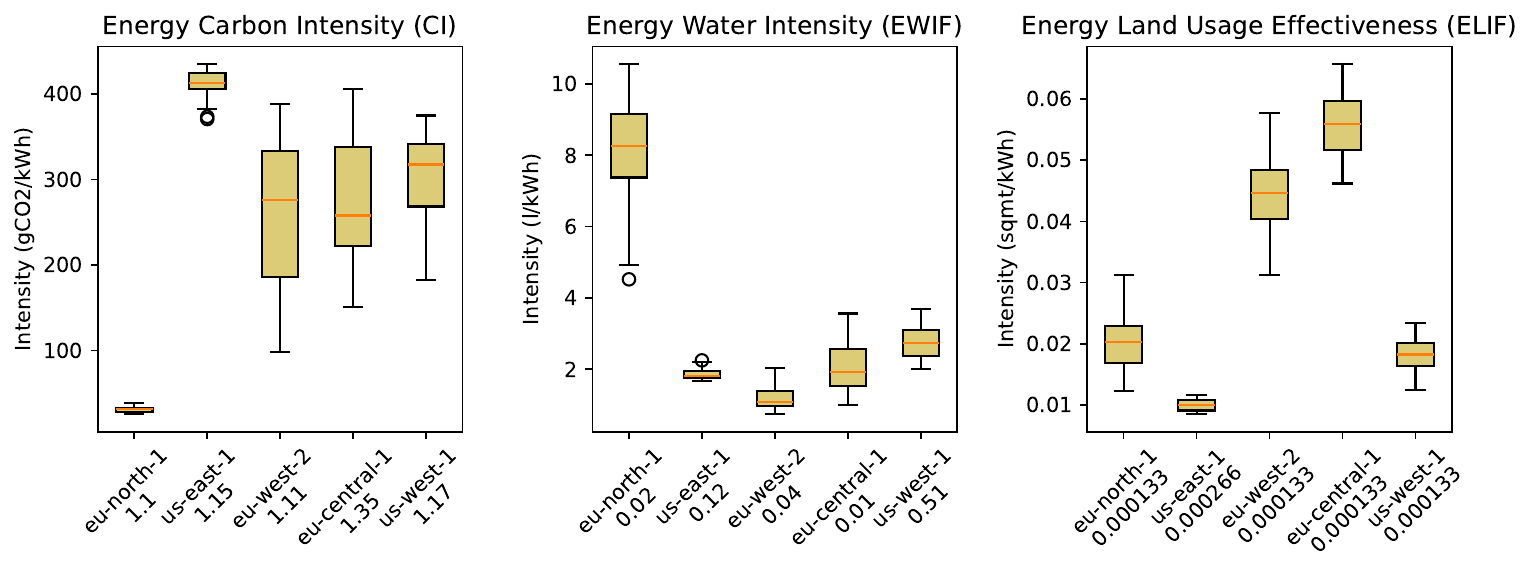}
    \caption{Grid intensities for each factor, and each region for Scenario 2 over a winter week, with an MAE of $10\%$.}
    \label{fig:aws_grid_intensities}
\end{figure*}

Figure~\ref{fig:aws_load} shows the request distribution across the regions.
To help us understand this distribution, let us consider Figure ~\ref{fig:aws_grid_intensities}, which simultaneously reports intensity values over the winter week and the efficiency metrics for each factor for each region.

When optimizing for carbon, about 50\% of requests are scheduled to "eu-north-1", which is the most energy-efficient region (PUE 1.1) and also the region with the grid with the lowest carbon intensity. 
When optimizing for water, about 40\% of requests are scheduled to "eu-west-2", which is not the most water efficient region (WUE 0.04), but it has the grid with the lowest impact on water.
Finally, when optimizing for land use, about 45\% of requests are scheduled to "us-east-1", which, even if it is the region with the most land occupation, it is the one where the grid has the lowest impact on land use.

\begin{figure*}
    \centering
    \includegraphics[width=\linewidth]{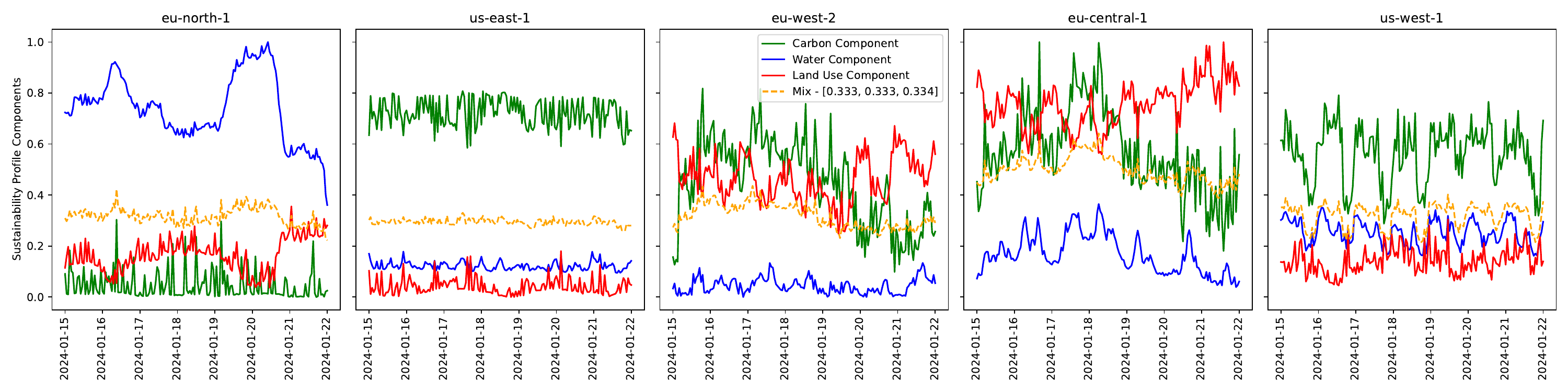}
    \caption{\review{Sustainability Profile components for Scenario 2 over a winter week, with an MAE of $10\%$.}}
    \label{fig:aws_sust_profiles}
\end{figure*}

\review{
Figure~\ref{fig:aws_sust_profiles} shows the sustainability profiles over time.
As in Scenario 1, the land use component of the sustainability profiles appears to dominate the others, albeit to a lesser extent. In fact, the average distance between the mix configuration signal and the components of the sustainability profile are 0.24, 0.25 and 0.20, respectively, for carbon, water and land use. 
Also for this scenario, this explains the similarity between the results of the land-only configuration and the mixed configuration.
}

\begin{figure*}
    \centering
    \includegraphics[width=\linewidth]{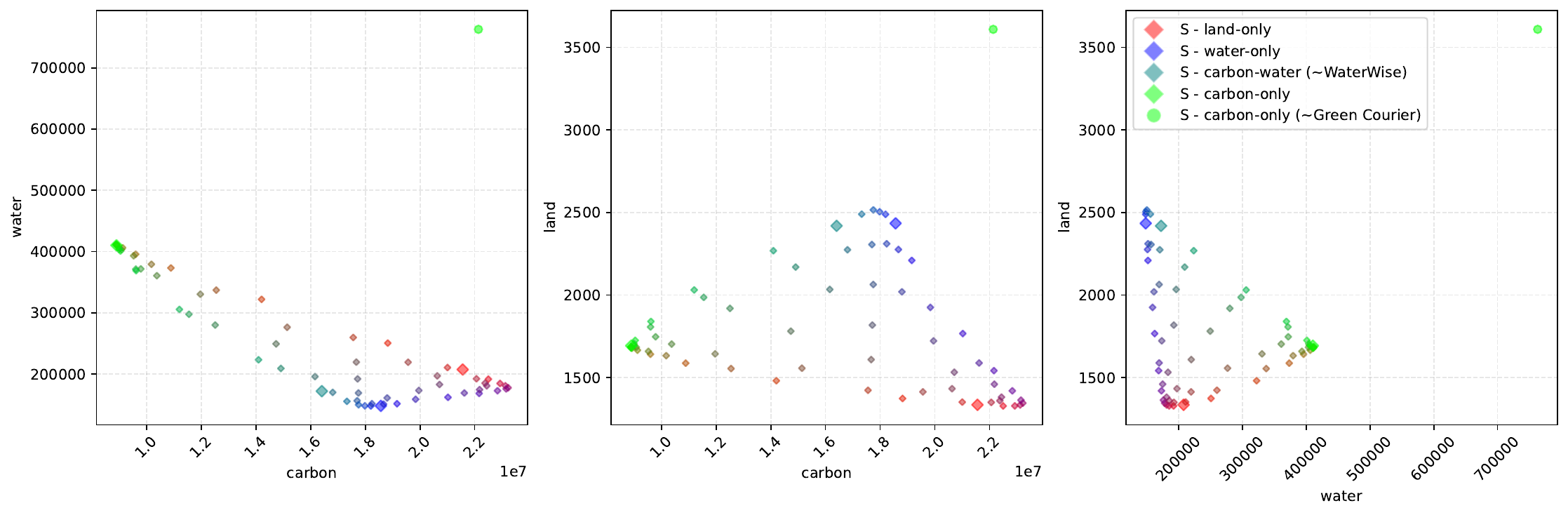}
    \caption{\review{2D - Pareto Analysis for Scenario 2 over a winter week, with an MAE of $10\%$.}}
    \label{fig:aws_pareto}
\end{figure*}

\review{
\paragraph{Pareto Analysis}
The same Pareto analysis was repeated for Scenario 2. Figure~\ref{fig:aws_pareto} shows the resulting points projected onto the three pairwise objective planes: carbon vs. water, carbon vs. land, and water vs. land. 
Four of the five representative configurations lie on the Pareto front and are therefore non-dominated. The exception is the carbon-only configuration ($\sim$GreenCourier), which is dominated, as expected from the results shown in Figure~\ref{fig:aws_winter_0.1_improvement_spatiotemp}. This analysis also confirms a genuine three-way trade-off in this scenario. The Pareto analysis of this scenario seems to confirm the correlation of water and land, as well as the fact that land is pretty much uncorrelated.
}


\paragraph{Migration Overheads of Spatial Shifting}
For this scenario, also an alternative approach to spatial shifting (S) \rew{is considered}. In this approach, the data associated with each request must be transferred each time a migration occurs, even if the data was transferred to that region previously.
Results show that this approach still achieves significant improvements over the baseline, but SP achieves about $5-6\%$ better results in the optimized metric.
\begin{table}[]
\centering

\begin{tabular}{|p{1.3cm}|cc|cc|cc|}
\hline
& \multicolumn{2}{c}{\textit{Carbon (\%)}}   & \multicolumn{2}{|c|}{\textit{Water (\%)}} & \multicolumn{2}{c|}{\textit{Land Use (\%)}} \\
                & S & SP   & S & SP   & S & SP  \\
\hline
carbon-only   & 12.67 & 6.96 & 3.45  & 1.80& 7.68 & 4.13 \\
water-only    & 3.50 & 1.85 & 5.47 & 2.94 & 3.07 & 1.58 \\
land-use-only & 3.51 & 1.80 & 4.55 & 2.60 & 6.62 & 3.65\\
mix           & 1.69 & 0.85 & 1.67 & 0.89 & 1.89 & 1.01 \\
\hline
\end{tabular}

\caption{Spatial shifting (persistent storage vs non-persistent storage) average percentage of migration footprints for Scenario 2 during Winter week, with an MAE of $10\%$.}\label{tab:migration_overhead}
\end{table}

Table~\ref{tab:migration_overhead} provides a summary of the migration overhead for the three factors; it shows the percentage of footprints caused by migration. The migration overhead varies based on the scheduling algorithm and the weights used, ranging from 0.85\% to 12.67\% for carbon, 0.89\% to 5.47\% for water, and 1.01\% to 7.68\% for land use. The overhead of SP is about half compared to S.

\review{
In addition to the environmental impact of migration itself, Figure~\ref{fig:aws_migrations} presents an analysis of the monetary costs of data transfer (see Appendix~\ref{online_res}). 

Results show that monetary migration cost varies substantially depending on the weighting used, from a median of approximately \$320 under the mix configuration to approximately \$1450 under the carbon-only configuration, roughly a $4$--$5\times$ difference. 
These results are shown alongside the percentage of cross-continental migrations. The land-use-only configuration, despite its relatively low monetary cost, produces the highest rate of violations (median $\sim0.86\%$), while water-only achieves the lowest ($\sim0.39\%$). 
The results show that the monetary costs are significant and that cross-continental migrations raise concerns, as they can expose workload data to different privacy regimes and weaken guarantees of data sovereignty. 
Moreover, no single weighting configuration is jointly optimal across environmental, economic, and regulatory dimensions.
However, modeling these risks and costs in depth is beyond the scope of this study and will be the subject of future research.
}
\begin{figure*}
    \centering
    \includegraphics[width=\linewidth]{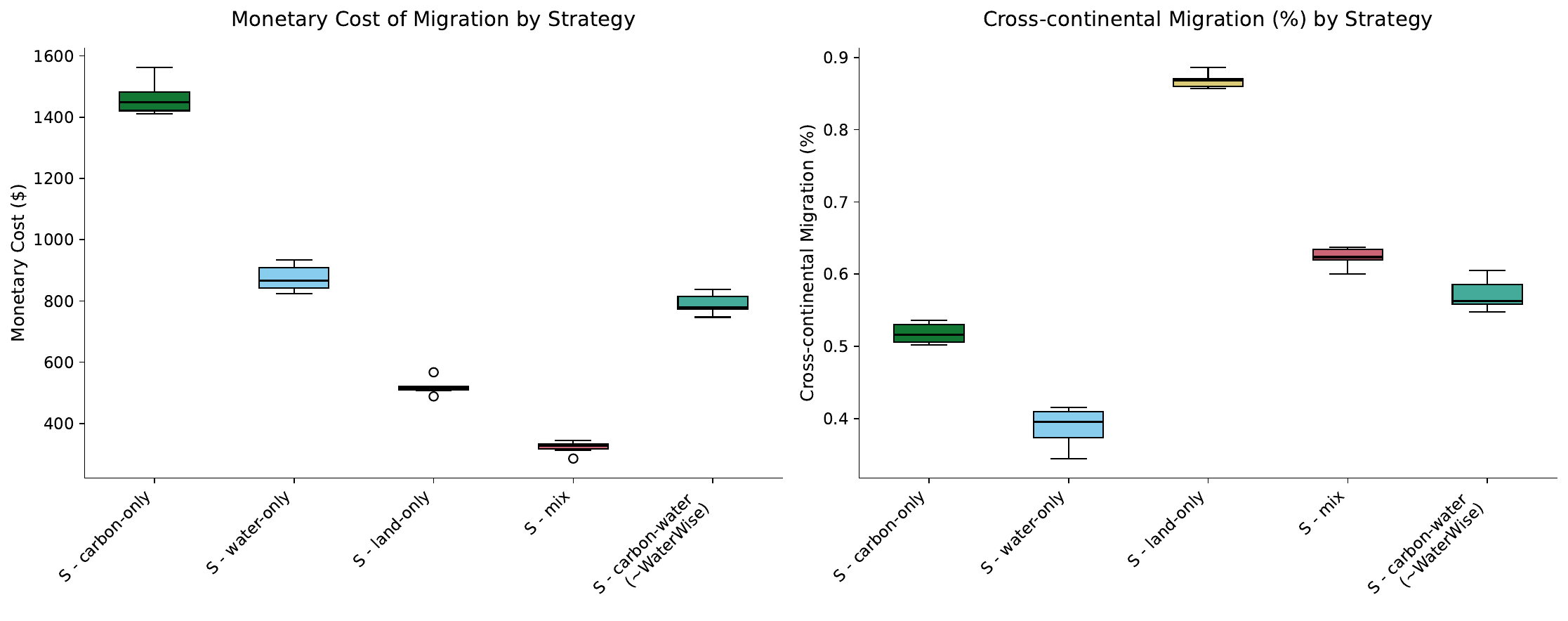}
    \caption{\review{Monetary cost (left) and cross-continental percentage (right) of migrations under the S strategy for Scenario 2 during the Winter week, broken down by scheduling weight configuration.}}
    \label{fig:aws_migrations}
\end{figure*}

\paragraph{Spatio-Temporal Shifting}
Finally, \review{as shown in Figure~\ref{fig:aws_winter_0.1_improvement_spatiotemp},} the best outcome is achieved by the spatio-temporal shifting approach (STP) with a delay tolerance of 4 hours, which results in 55\% improvement in carbon footprint when optimizing for carbon, about 40\% improvement in water footprint when optimizing for water, and about 48\% improvement in land use footprint when optimizing for land use.
However, these improvements with respect to only spatial shifting are small.

\review{
This can be further explained by looking at Figure\ref{fig:aws_sust_profiles}. It can be seen that fluctuations over time within a single region are limited for each component or linear combination of components. This limits what can be achieved by exploiting only temporal shifting. Meanwhile, the signal differences between regions are more evident, allowing greater footprint savings. Combining the two strategies enables both the more evident differences between regions and the intra-day fluctuation of the profile components to be exploited.
}

\paragraph{Sensitivity Analysis}
Following the same approach used for Scenario 1, sensitivity analysis for this scenario \rew{was conducted} as well. 
Tables~\ref{tab:aws_sa_season} and~\ref{tab:aws_sa_mix} respectively show the footprint improvements over the local baseline for different seasons and for different prediction errors for the grid renewable share (mae).
The results fluctuate slightly with the prediction error (2-3\%), but the improvements remain significant even with a high prediction error of 0.2. Also, the season does not affect the results significantly, with variations of 4\% at most, considering the optimized factor.

\section{Discussion}\label{sec:discussion}
In this section, the results obtained \rew{are discussed}. 

\subsection{Scenario 1: Azure and FaaS}
The results demonstrate that regional shifting of FaaS workloads across Azure regions leads to noticeable improvements in carbon efficiency.

Overall, these results highlight the potential of spatial shifting to significantly enhance the sustainability of FaaS workloads by leveraging regional variations in grid intensities and regional efficiency differences.
This suggests that lightweight workloads such as FaaS functions could serve as a first mover for sustainability-aware orchestration, as they can be shifted freely without incurring significant migration penalties. Our results show that accounting for provider-declared PUE/WUE improves allocation decisions, especially when grid mixes are similar. This highlights that relying solely on grid intensity overlooks potential sustainability gains and the importance of region-specific data center disclosures for enabling sustainability-aware workload management.


\subsection{Scenario 2: AWS and Big Data}\label{sec:discussion_aws}

The results highlight the potential of spatial shifting to significantly enhance the sustainability of cloud workloads by leveraging, for the most part, regional variations in grid intensities. 
However, the migration overhead is relevant, especially when data is not persistently stored after migration. Nevertheless, the benefits of spatial shifting in reducing the overall footprint outweigh the footprints associated with migration, obtaining a substantial footprint improvement.
Combined spatial and temporal shifting can lead to even more significant improvements, although the additional benefits are relatively modest compared to spatial shifting alone.
This suggests that while temporal shifting can contribute to sustainability goals, its impact is limited by the inherent variability of renewable energy availability within a single region.
Though it is worth noting that even if temporal shifting has less savings, it does not require to move data between regions, which sometimes will not be possible due to security/privacy concerns.

These findings reveal a trade-off: although spatial shifting yields the greatest reductions, it may be limited in practice by sovereignty, privacy, and cost constraints. Therefore, temporal shifting is an important alternative, despite its standalone benefits being smaller; it can be applied even when workloads cannot be transferred to another region. This suggests prioritizing spatial shifting for portable workloads and using temporal shifting as a fallback for constrained ones.

\subsection{Comparison of the Two Scenarios}
With diverse providers and workloads, the results are different, but the improvements are relevant in both. 
In particular, for the Azure-based scenario the greatest improvements \rew{were obtained}, since there is no footprint overhead when migrating, while for the AWS-based scenario, there can be relevant overhead, e.g., 12\% carbon footprint related to data transfer. However, this overhead is such that the improvements are still significant. Moreover, it allows us to obtain a more balanced distribution of requests among the regions.

\subsection{Trade-Off Analysis}
As expected, in both scenarios, optimizing for one metric can lead to trade-offs in the other metrics. For instance, optimizing for carbon footprint leads to a significant increase in water footprint, as already observed in previous studies~\cite{jiang2025waterwise}.

This is particularly evident when optimizing for carbon, which results in a significant increase in water footprint. Optimizing for water results in an increase in both carbon and land use footprints, but to a lesser extent compared to the increase in water footprint when optimizing only for carbon. Optimizing for the mix of the three factors results in a balanced reduction of all three footprints, but the improvements are lower compared to optimizing for only one factor.

\begin{figure*}
    \centering
    \includegraphics[width=\linewidth]{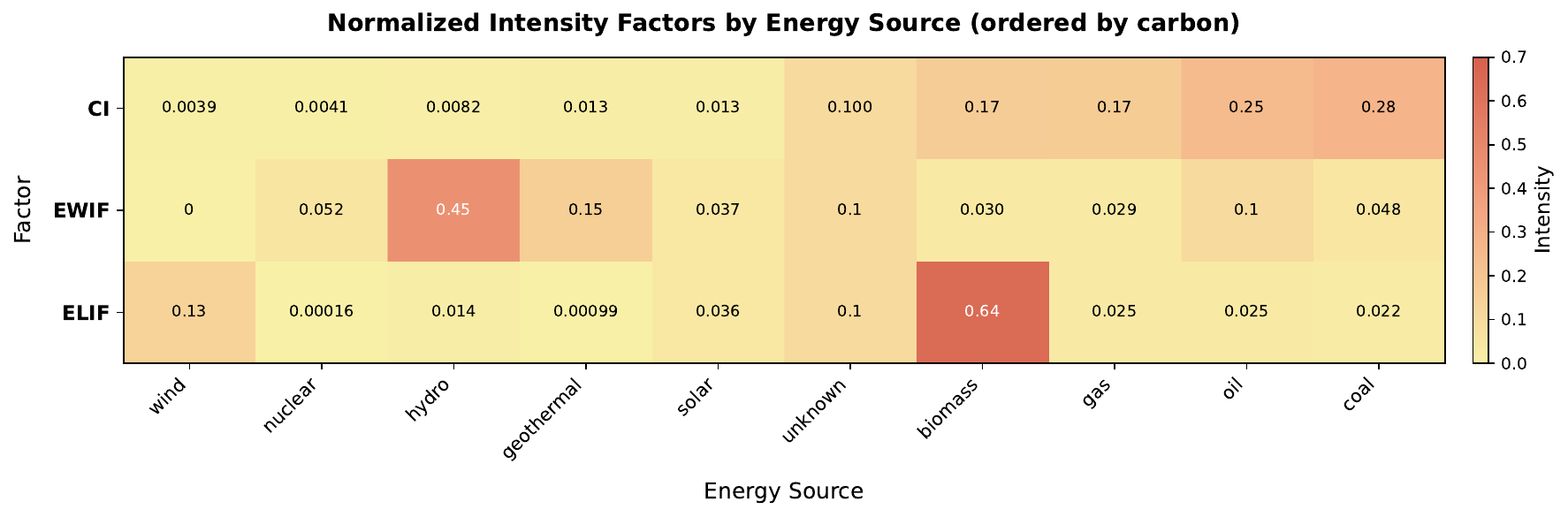}
    \caption{Normalized energy intensity factors for each energy source, showing a negative correlation between carbon and water, and seemingly no correlation between land and the other factors.}
    \label{fig:eif}
\end{figure*}

To understand these trade-offs, consider the energy intensity factors for each energy source.
Figure~\ref{fig:eif} shows the energy intensity factors normalized between 0 and 1 (by dividing by the sum of each factor across all sources).
It seems that carbon and water are negatively correlated (e.g., hydro, coal), but not all low-carbon-intensity sources have high water intensity (e.g., wind and nuclear).
On the other hand, land use seems to not be correlated with the other factors, although there are sources with high land use and low water intensity (e.g. biomass), and sources with low land use and high carbon and water intensity (e.g. oil).

\subsection{Threats to Validity}

This study presents some threats to validity that should be considered when interpreting the results. 

\review{
\paragraph{Simulation-Based Methodology} This study relies on simulation rather than deployment on production infrastructure. 
}
\rew{It is assumed} that there are always resources available, which might not be true in reality. However, considering the promise of elastic scalability of public cloud providers and general availability of surplus ``spot'' resources, \review{specifically for the large-scale, elastic, hyperscale infrastructures operated by major providers such as AWS and Azure,} 
this appears to be a sensible assumption.
\review{This is supported by empirical measurement studies of spot instance markets, such as SpotLake~\cite{9975369}, which continuously track spot instance availability, interruption rates, and pricing across major cloud vendors and show that surplus capacity is persistently made available across regions.}
\review{
More broadly, the simulations are grounded in real-world infrastructure characteristics and workload traces to the extent allowed by public data availability.

The choice of a simulation-based approach is also supported by the fact that it allows  to study a large number of geographically distributed scenarios and scheduling weight configurations without incurring the substantial computational cost of actually executing every evaluated workload and configuration in production.
}

\paragraph{Generality} The scenarios considered may not fully represent the diversity of real-world cloud environments, as \rew{this work is} focused on specific providers (Azure and AWS) and workload types (FaaS and big data). Moreover, the workload traces used for the AWS-based scenario may not be representative of production cloud workloads since arrival patterns are synthetically randomized. However, selected two diverse cases \rew{were selected}: lightweight, latency-sensitive workloads (FaaS) and scalable data-intensive batch processing (big data). The fact that we observe similar trade-offs across such different cases increases the confidence that the results generalize, even though more workload classes should be studied in future work.
\review{Our conclusions may be questioned on the basis that they are valid only for the specific regions and grid mixes considered, and not for other regions or grid mixes. 
However, as it is the grid mix that dominates the shifting decisions, it can be argued that the observed trade-offs follow from the physical characteristics of electricity generation technologies. These characteristics are intrinsic to the technologies themselves, rather than being specific to any particular region.
Therefore, the qualitative trade-off patterns (e.g. the carbon–water trade-off) are expected to be observed also in other regions and providers with a similar generation technology mix, although the exact scale of the trade-off will vary according to the local mix. Validating this transferability across additional regions, providers, and datasets is an important area for future research.}

\paragraph{Input Data Quality} The accuracy of our results is contingent upon the quality of the input data, including grid mix predictions. In our case, these predictions were mimicked using historical data and applying an arbitrary noise distribution. 
\review{This approach is consistent with other carbon-aware scheduling practices, in which noise is applied to carbon intensity signals~\cite{10.1145/3464298.3493399, 11120561}. Moreover, this technique made it possible to test our approach at different magnitudes of error, as demonstrated by the sensitivity analysis.  However, we recognize that repeating the experiments using an actual operational forecasting model would be an important step for future research.  }
Moreover, to estimate the LUE parameter, estimations for the region's IT power consumption and the land occupation \rew{were used}, since precise data is not available.
\review{
Note that, unlike PUE and WUE, LUE has no established reporting convention. Cloud providers do not disclose site-specific IT power consumption alongside facility land area. Consequently, our computed LUE values cannot be directly validated against figures disclosed by providers. This is mitigated by grounding our land occupation estimates in publicly reported real-estate and regulatory sources (e.g., news reporting on data center land acquisitions, and AWS's Form 10-K filings) rather than assumed or synthetic values.
}
\review{
Further note that data center operators would have access to precise, site-specific land occupation and IT power consumption figures. Thus, our reliance on estimation reflects the limits of publicly available data and not the inherent unavailability of such measurements.
}
These approximations could affect the validity of our findings, yet this data is unfortunately not freely and openly available.
Finally, it is assumed that the application runtimes are available to the schedulers. While it is not necessarily realistic to know these runtimes exactly prior to executions, low-error estimates could be available from previous runs or profiling~\cite{10.14778/2556549.2556553, 9610321, 10.1145/3514221.3517892, 194946}.
\review{
\paragraph{Embodied Impacts} As discussed in Section~\ref{sec:background}, this study only considers the operational share of carbon, water and land footprints. Embodied impacts arising from the manufacture of IT equipment and the construction of data center facilities are excluded.  However, it would be valuable to extend our multi-objective formulation to incorporate these impacts in future work.
}

\review{
\paragraph{Regional Resource Equivalence} 
Both our land use metrics ($LUE_d$, $ELIF$) and our water footprint metrics ($WUE_d$, $EWIF$) treat a unit of the respective resource as equivalent, regardless of region. One square metre of land and one litre of water are given the same weight, whether they are consumed in an area with an abundance or shortage of water, or on ecologically valuable or already degraded land. However, the environmental significance of a given quantity of water or land consumption depends heavily on regional context. Incorporating such regional weighting for water and land use would be valuable for future studies.
}
\paragraph{Spatial Shifting Constraints} Our models assume that workloads can be shifted across regions and time without considering potential constraints such as latency requirements, data sovereignty issues, or privacy concerns.
\review{Our results show that the current optimization setting can lead to a high percentage of cross-continental migrations, which would need to be reconciled with such constraints if they were to be enforced. A more thorough treatment of these constraints, including their integration into the scheduling objective itself, is left for future work.}
While these issues are significant, they primarily affect spatial shifting. Temporal shifting, in contrast, only delays the execution, and all data can remain in the same location.

\paragraph{Economic Costs} Our simulations do not \review{explicitly optimize for the} economic costs associated with shifting workloads geographically.
In practice, migrating data between data center locations can introduce significant costs,\review{ and our results show that these costs vary substantially across scheduling configurations}, which would need to be considered \review{when selecting a deployment configuration in practice, alongside the associated jurisdiction and data-sovereignty implications discussed above}.
Meanwhile, \rew{it is assumed that} the cost of storing data for longer for delaying executions has a negligible monetary cost.


\section{Conclusion}
\label{sec:conclusions}

In this paper, \rew{has been} demonstrated that spatial and temporal shifting can significantly reduce the environmental footprint of cloud workloads in terms of carbon emissions, water consumption, and land use. Spatial shifting, in particular, has yielded substantial improvements across all three environmental factors considered, with reductions ranging from 20\% to 85\% depending on the scenario and optimization criteria. Temporal shifting also contributes to footprint reduction, albeit to a lesser extent, with improvements of up to 12\%. However, since temporal shifting does not require workloads to be migrated between regions, it may be more relevant in practice, when migration is restricted for data privacy or cost reasons. The combination of both methods yields the highest reductions, although the incremental benefit over spatial shifting alone is relatively small. 

Our sensitivity analyzes indicate that both shifting methods are robust to prediction errors in grid mix data and maintain their effectiveness across different seasons. 
Finally, \rew{has been} demonstrated that it is possible to reduce the environmental footprint and mitigate trade-off implications by optimizing multiple impact factors simultaneously using a mixed strategy.
Overall, our findings highlight the potential of a more sustainable workload scheduling as a practical approach considering diverse cloud providers and workloads.

In the future, the impact of cross-region availability, load balancing, and low-latency requirements \rew{should be studied}. In addition, beyond shifting, the role of resource configurations and dynamic scaling when carbon, water, and land use are jointly optimized for cloud workloads \rew{should be further explored}.
Finally, it would be interesting to incorporate the concept of ``water stress'' to reflect the varying impact that water consumption can have on areas with different hydrogeological characteristics to provide a more equitable representation of the environmental trade-off across regions. Similarly, a concept of ``land-use stress'' could be introduced to better reflect the environmental impact of data center developments in different geographical areas.

\section*{Acknowledgments}
This work was supported by the Engineering and Physical Sciences Research Council under grant number UKRI154 (“Casper: Carbon-Aware Scalable Processing in Elastic Clusters”).

\section*{Rights Retention}

For the purpose of open access, we have applied a Creative Commons Attribution (CC BY) licence to any Author Accepted Manuscript version arising from this submission.

{
\begin{table*}[]
    \centering
    \begin{tabular}{c||l|rrrrrr}
 \multirow{2}{*}{algorithm} & \multirow{2}{*}{period} & \multicolumn{2}{r}{\makecell{carbon\\improvement}} & \multicolumn{2}{r}{\makecell{water\\improvement}} & \multicolumn{2}{r}{\makecell{land use\\improvement}}\\
& & mean & std & mean & std & mean & std  \\
\hline
\multirow[t]{4}{*}{S - carbon-only} & autumn & 91.28 & 0.03 & -118.26 & 0.75 & -8.10 & 0.46 \\
 & spring & 86.56 & 0.04 & -139.90 & 1.17 & 18.29 & 0.32   \\
 & summer & 91.87 & 0.02 & -123.28 & 1.15 & -13.85 & 0.29 \\
 & winter & 90.55 & 0.03 & -129.41 & 1.12 & -15.46 & 0.26  \\
\hline

\multirow[t]{4}{*}{S - water-only} & autumn & -48.74 & 0.55 & 55.33 & 0.22 & 6.57 & 0.35  \\
 & spring & -29.97 & 0.39 & 59.40 & 0.19 & -11.63 & 0.41 \\
 & summer & -24.66 & 0.39 & 50.82 & 0.21 & 32.46 & 0.13  \\
 & winter & -42.75 & 0.52 & 56.42 & 0.23 & 5.99 & 0.32 \\
\hline
\multirow[t]{4}{*}{S - land-only} & autumn & -25.17 & 0.50 & 40.90 & 0.30 & 57.00 & 0.17 \\
 & spring & -15.63 & 0.35 & 34.32 & 0.37 & 42.24 & 0.22\\
 & summer & -13.88 & 0.41 & 39.31 & 0.36 & 37.54 & 0.12\\
 & winter & -23.12 & 0.42 & 46.16 & 0.27 & 45.40 & 0.19 \\
\hline

\multirow[t]{4}{*}{S - mix} & autumn & -25.86 & 0.49 & 42.35 & 0.27 & 57.16 & 0.17  \\
 & spring & -23.35 & 0.35 & 49.22 & 0.24 & 42.73 & 0.22 \\
 & summer & -17.64 & 0.40 & 49.55 & 0.23 & 38.33 & 0.12  \\
 & winter & -23.34 & 0.43 & 47.10 & 0.28 & 45.74 & 0.18 \\
\hline
\end{tabular}
\caption{Scenario 1 - Season sensitivity analysis}

\label{tab:azure_sa_season}
\end{table*}
}

{
\begin{table*}[]
    \centering

    \begin{tabular}{c||l|rrrrrr}
 \multirow{2}{*}{algorithm} &  \multirow{2}{*}{mae} & \multicolumn{2}{r}{\makecell{carbon\\improvement}} & \multicolumn{2}{r}{\makecell{water\\improvement}} & \multicolumn{2}{r}{\makecell{land use\\improvement}} \\
& & mean & std & mean & std & mean & std \\
\hline
\multirow[t]{4}{*}{S - carbon-only} &  0.05 & 90.55 & 0.03 & -129.41 & 1.12 & -15.46 & 0.26 \\
 & 0.1 & 90.55 & 0.03 & -129.41 & 1.12 & -15.46 & 0.26 \\
 & 0.15 & 90.55 & 0.03 & -129.41 & 1.12 & -15.46 & 0.26 \\
 & 0.2 & 90.55 & 0.03 & -129.41 & 1.12 & -15.46 & 0.26 \\
\hline

\multirow[t]{4}{*}{S - water-only} & 0.05 & -43.64 & 0.52 & 56.60 & 0.23 & 4.95 & 0.32 \\
 & 0.1 & -42.75 & 0.52 & 56.42 & 0.23 & 5.99 & 0.32 \\
 & 0.15& -42.81 & 0.51 & 56.25 & 0.23 & 6.79 & 0.32 \\
 & 0.2 & -41.36 & 0.51 & 55.84 & 0.23 & 8.59 & 0.30 \\
\hline
\multirow[t]{4}{*}{S - land-only}  & 0.05 & -23.40 & 0.43 & 47.12 & 0.28 & 45.71 & 0.18 \\
 & 0.1 & -23.12 & 0.42 & 46.16 & 0.27 & 45.40 & 0.19 \\
 & 0.15& -22.33 & 0.42 & 44.14 & 0.29 & 44.85 & 0.18 \\
 & 0.2 & -22.03 & 0.42 & 42.84 & 0.30 & 44.15 & 0.19 \\
\hline

\multirow[t]{4}{*}{S - mix} & 0.05 & -23.34 & 0.43 & 47.10 & 0.28 & 45.74 & 0.18 \\
 & 0.1 & -23.34 & 0.43 & 47.10 & 0.28 & 45.74 & 0.18 \\
 & 0.15& -23.34 & 0.43 & 47.10 & 0.28 & 45.74 & 0.18 \\
 & 0.2 & -23.32 & 0.43 & 47.08 & 0.28 & 45.57 & 0.18 \\
\hline
\end{tabular}
\caption{Scenario 1 - Mix renewable share prediction error sensitivity analysis}
\label{tab:azure_sa_mix}
\end{table*}
}

{
\begin{table*}[]
    \centering

\begin{tabular}{c||l|rrrrrr}
 \multirow{2}{*}{algorithm} & \multirow{2}{*}{period} & \multicolumn{2}{r}{\makecell{carbon\\improvement}} & \multicolumn{2}{r}{\makecell{water\\improvement}} & \multicolumn{2}{r}{\makecell{land use\\improvement}}  \\
& & mean & std & mean & std & mean & std  \\
\hline
\multirow[c]{4}{*}{\makecell{SP\\carbon-only}} & autumn & 52.00 & 1.33 & -65.81 & 2.96 & 21.26 & 0.86 \\
 & spring & 45.96 & 1.11 & -68.89 & 2.95 & 30.19 & 0.89 \\
 & summer & 51.05 & 1.34 & -70.26 & 3.13 & 29.75 & 0.90  \\
 & winter & 51.60 & 1.32 & -88.18 & 3.86 & 20.05 & 1.00  \\
\hline
\multirow[c]{4}{*}{\makecell{SP\\water-only}} & autumn & -3.02 & 0.88 & 32.83 & 1.16 & -18.67 & 1.57 \\
 & spring & -0.53 & 0.81 & 32.77 & 1.32 & -20.88 & 1.64 \\
 & summer & -27.04 & 2.13 & 29.59 & 1.20 & 9.16 & 1.38 \\
 & winter & -3.81 & 0.96 & 34.42 & 1.46 & -19.40 & 1.72  \\
\hline

\multirow[c]{4}{*}{\makecell{SP\\land-only}} & autumn & -25.84 & 2.05 & 11.83 & 1.63 & 41.04 & 1.02  \\
 & spring & -21.40 & 1.81 & -3.26 & 1.57 & 38.43 & 0.98  \\
 & summer & -23.65 & 1.46 & 2.72 & 1.45 & 39.74 & 1.00 \\
 & winter & -27.72 & 1.81 & 11.42 & 1.76 & 37.49 & 1.00 \\
\hline

\multirow[c]{4}{*}{\makecell{SP\\mix}} & autumn & 1.99 & 0.90 & 7.43 & 1.13 & 24.13 & 0.78\\
 & spring & 2.78 & 1.24 & 15.98 & 1.67 & 16.26 & 1.20 \\
 & summer & 6.49 & 0.92 & -0.22 & 1.33 & 20.48 & 1.08   \\
 & winter & -0.59 & 0.64 & 3.33 & 1.13 & 17.39 & 0.99 \\
\hline

\multirow[c]{4}{*}{\makecell{T (dt4)\\carbon-only}} & autumn & 5.09 & 0.26 & 3.51 & 0.10 & 0.59 & 0.09 \\
 & spring & 4.28 & 0.28 & 2.63 & 0.16 & 1.37 & 0.11  \\
 & summer & 4.60 & 0.23 & 2.00 & 0.12 & 1.46 & 0.17  \\
 & winter & 2.02 & 0.14 & 1.90 & 0.08 & -0.79 & 0.03  \\

\hline
 \multirow[c]{4}{*}{\makecell{T (dt4)\\water-only}} & autumn & 4.93 & 0.24 & 6.81 & 0.09 & 0.32 & 0.11 \\
 & spring & 4.62 & 0.27 & 6.24 & 0.16 & 0.51 & 0.10  \\
 & summer & 4.35 & 0.24 & 7.26 & 0.16 & 0.99 & 0.13 \\
 & winter & 1.91 & 0.11 & 4.49 & 0.11 & -1.33 & 0.05  \\

\hline
\multirow[c]{4}{*}{\makecell{T (dt4)\\land-only}} & autumn & 1.97 & 0.23 & -0.15 & 0.13 & 3.71 & 0.12  \\
 & spring & 2.67 & 0.24 & 0.48 & 0.17 & 3.75 & 0.14  \\
 & summer & 2.11 & 0.20 & -0.21 & 0.09 & 4.80 & 0.20  \\
 & winter & -0.65 & 0.04 & -1.41 & 0.05 & 1.62 & 0.05 \\
\hline

\multirow[c]{4}{*}{\makecell{T (dt4)\\mix}} & autumn & 5.54 & 0.25 & 5.49 & 0.09 & 1.71 & 0.13  \\
 & spring & 5.67 & 0.29 & 4.87 & 0.18 & 2.74 & 0.12 \\
 & summer & 5.24 & 0.26 & 5.77 & 0.15 & 3.12 & 0.17  \\
 & winter & 2.22 & 0.13 & 3.19 & 0.11 & -0.10 & 0.03 \\
\hline

\multirow[c]{4}{*}{\makecell{SPT (dt4)\\carbon-only}} & autumn & 54.46 & 1.30 & -63.54 & 3.00 & 21.24 & 0.87 \\
 & spring & 48.67 & 1.13 & -72.15 & 3.29 & 31.87 & 0.91 \\
 & summer & 53.37 & 1.34 & -71.52 & 3.22 & 30.97 & 0.93  \\
 & winter & 52.59 & 1.31 & -86.23 & 3.87 & 19.37 & 1.02 \\
\hline

\multirow[c]{4}{*}{\makecell{SPT (dt4)\\water-only}} & autumn & 1.66 & 0.83 & 37.05 & 1.13 & -19.98 & 1.42  \\
 & spring & 3.93 & 0.73 & 36.43 & 1.31 & -23.77 & 1.63  \\
 & summer & -22.21 & 1.75 & 33.30 & 1.18 & 6.84 & 1.07  \\
 & winter & -1.79 & 0.93 & 37.66 & 1.38 & -21.60 & 1.72   \\

\hline

\multirow[c]{4}{*}{\makecell{SPT (dt4)\\land-only}} & autumn & -23.46 & 2.02 & 10.23 & 1.51 & 42.53 & 0.99 \\
 & spring & -22.95 & 2.00 & 0.64 & 1.34 & 40.20 & 0.96  \\
 & summer & -22.05 & 1.57 & 2.83 & 1.15 & 41.86 & 1.00  \\
 & winter & -28.16 & 1.93 & 11.35 & 2.01 & 38.11 & 1.01  \\
\hline

\multirow[c]{4}{*}{\makecell{SPT (dt4)\\mix}} & autumn & 9.21 & 1.09 & 10.30 & 1.35 & 24.10 & 0.82 \\
 & spring & 12.29 & 0.55 & 18.50 & 1.66 & 17.32 & 1.09 \\
 & summer & 12.86 & 0.78 & 3.76 & 1.42 & 20.38 & 0.87  \\
 & winter & 3.27 & 0.49 & 5.09 & 0.88 & 16.91 & 1.17 \\
\hline
\end{tabular}
\caption{Scenario 2 - Season sensitivity analysis }
\label{tab:aws_sa_season}
\end{table*}}

{
\begin{table*}[]
    \centering

\begin{tabular}{c||l|rrrrrr}
 \multirow{2}{*}{algorithm} & \multirow{2}{*}{mae} & \multicolumn{2}{r}{\makecell{carbon\\improvement}} & \multicolumn{2}{r}{\makecell{water\\improvement}} & \multicolumn{2}{r}{\makecell{land use\\improvement}} \\
& & mean & std & mean & std & mean & std \\
\hline
\multirow[c]{4}{*}{\makecell{SP\\carbon-only}} & 0.05 & 51.72 & 1.32 & -88.67 & 3.86 & 20.31 & 0.97 \\
& 0.1 & 51.60 & 1.32 & -88.18 & 3.86 & 20.05 & 1.00 \\
& 0.15& 51.27 & 1.34 & -86.49 & 3.91 & 19.29 & 0.99 \\
& 0.2& 49.97 & 1.25 & -81.37 & 3.67 & 16.97 & 0.95 \\
\hline
\multirow[c]{4}{*}{\makecell{SP\\water-only}} & 0.05& -3.61 & 0.95 & 34.71 & 1.46 & -20.22 & 1.65 \\
&0.1& -3.81 & 0.96 & 34.42 & 1.46 & -19.40 & 1.72 \\
&0.15& -4.27 & 1.00 & 34.16 & 1.44 & -19.20 & 1.71 \\
&0.2& -4.04 & 1.04 & 34.17 & 1.44 & -18.63 & 1.71 \\
\hline

\multirow[c]{4}{*}{\makecell{SP\\land-only}} & 0.05& -28.87 & 1.87 & 12.51 & 1.87 & 37.73 & 1.01 \\
 & 0.1& -27.72 & 1.81 & 11.42 & 1.76 & 37.49 & 1.00 \\
 & 0.15& -27.32 & 1.71 & 10.60 & 1.63 & 37.32 & 1.01 \\
 & 0.2& -25.64 & 1.57 & 8.49 & 1.71 & 36.78 & 0.95 \\
\hline

\multirow[c]{4}{*}{\makecell{SP\\mix}} & 0.05& 0.17 & 0.56 & 1.58 & 1.03 & 17.69 & 1.20 \\
 & 0.1& -0.59 & 0.64 & 3.33 & 1.13 & 17.39 & 0.99 \\
 & 0.15& 0.48 & 0.81 & 2.75 & 1.09 & 16.43 & 1.40 \\
 & 0.2& -6.59 & 1.11 & 8.37 & 1.43 & 19.93 & 1.29 \\
\hline

\multirow[c]{4}{*}{\makecell{T (dt4)\\carbon-only}} & 0.05& 2.67 & 0.15 & 1.55 & 0.07 & -0.89 & 0.04 \\
 & 0.1& 2.02 & 0.14 & 1.90 & 0.08 & -0.79 & 0.03 \\
 & 0.15& 1.92 & 0.13 & 1.39 & 0.09 & -0.62 & 0.03 \\
 & 0.2& 1.72 & 0.11 & 1.04 & 0.06 & -0.47 & 0.04 \\

\hline
 \multirow[c]{4}{*}{\makecell{T (dt4)\\water-only}} & 0.05& 2.15 & 0.12 & 4.85 & 0.11 & -1.60 & 0.06 \\
 & 0.1& 1.91 & 0.11 & 4.49 & 0.11 & -1.33 & 0.05 \\
 & 0.15& 1.76 & 0.11 & 4.34 & 0.11 & -1.29 & 0.06 \\
 & 0.2& 1.84 & 0.11 & 4.11 & 0.11 & -1.19 & 0.05 \\

\hline
\multirow[c]{4}{*}{\makecell{T (dt4)\\land-only}} & 0.05& -0.48 & 0.04 & -1.05 & 0.06 & 2.10 & 0.05 \\
 & 0.1& -0.65 & 0.04 & -1.41 & 0.05 & 1.62 & 0.05 \\
 & 0.15& -0.53 & 0.05 & -0.68 & 0.05 & 1.40 & 0.05 \\
 & 0.2& -0.08 & 0.07 & -0.51 & 0.05 & 1.22 & 0.05 \\
\hline

\multirow[c]{4}{*}{\makecell{T (dt4)\\mix}} & 0.05& 2.50 & 0.13 & 3.49 & 0.11 & -0.07 & 0.05 \\
 & 0.1& 2.22 & 0.13 & 3.19 & 0.11 & -0.10 & 0.03 \\
 & 0.15& 2.18 & 0.13 & 2.47 & 0.11 & -0.14 & 0.04 \\
 & 0.2& 1.95 & 0.12 & 2.43 & 0.10 & 0.00 & 0.04 \\
\hline

\multirow[c]{4}{*}{\makecell{SPT (dt4)\\carbon-only}} & 0.05& 52.86 & 1.31 & -86.21 & 3.86 & 19.38 & 1.02 \\
 & 0.1& 52.59 & 1.31 & -86.23 & 3.87 & 19.37 & 1.02 \\
 & 0.15& 52.52 & 1.32 & -86.48 & 3.87 & 19.46 & 1.03 \\
 & 0.2& 52.24 & 1.32 & -86.26 & 3.87 & 19.30 & 1.02 \\
\hline

\multirow[c]{4}{*}{\makecell{SPT (dt4)\\water-only}} & 0.05& -1.09 & 0.87 & 38.27 & 1.41 & -23.27 & 1.74 \\
 & 0.1& -1.79 & 0.93 & 37.66 & 1.38 & -21.60 & 1.72 \\
 & 0.15& -2.58 & 0.89 & 37.30 & 1.42 & -21.25 & 1.69 \\
 & 0.2& -2.06 & 0.93 & 37.11 & 1.45 & -20.60 & 1.79 \\

\hline

\multirow[c]{4}{*}{\makecell{SPT (dt4)\\land-only}} & 0.05& -28.82 & 1.84 & 12.47 & 1.87 & 38.60 & 0.97 \\
 & 0.1& -28.16 & 1.93 & 11.35 & 2.01 & 38.11 & 1.01 \\
 & 0.15& -28.63 & 1.93 & 12.43 & 1.94 & 38.17 & 1.01 \\
 & 0.2& -27.82 & 1.85 & 11.59 & 1.84 & 38.08 & 1.02 \\
\hline

\multirow[c]{4}{*}{\makecell{SPT (dt4)\\mix}} & 0.05& 2.88 & 0.46 & 4.36 & 0.76 & 18.20 & 1.13 \\
 & 0.1& 3.27 & 0.49 & 5.09 & 0.88 & 16.91 & 1.17 \\
 & 0.15& 5.33 & 0.59 & 3.57 & 0.85 & 15.25 & 1.09 \\
 & 0.2& 0.60 & 0.56 & 7.95 & 1.31 & 16.76 & 1.20 \\
\hline
\end{tabular}
\caption{Scenario 2 - Mix renewable share prediction error sensitivity analysis}

\label{tab:aws_sa_mix}
\end{table*}}


\bibliographystyle{elsarticle-num}
\bibliography{references}

@misc{ipcc2014_ar5_annexIII,
  title={Climate Change 2014: Mitigation of Climate Change. Annex III: Technology-specific cost and performance parameters},
  author={{IPCC}},
  year={2014},
}

@misc{nrel2011_water,
  title={Consumptive Water Use for U.S. Power Production},
  author={Macknick, Jordan and Newmark, Robin and Heath, Garvin and Hallett, KC},
  year={2011},
  institution={NREL},
}

@article{10.1145/3054177,
author = {Weerasiri, Denis and Barukh, Moshe Chai and Benatallah, Boualem and Sheng, Quan Z. and Ranjan, Rajiv},
title = {A Taxonomy and Survey of Cloud Resource Orchestration Techniques},
year = {2017},
issue_date = {March 2018},
publisher = {ACM},
address = {New York, NY, USA},
volume = {50},
number = {2},
issn = {0360-0300},
journal = {ACM Comput. Surv.},
month = {may},
articleno = {26}
}

@inproceedings{jiang2025waterwise,
author = {Jiang, Yankai and Roy, Rohan Basu and Kanakagiri, Raghavendra and Tiwari, Devesh},
title = {WaterWise: Co-optimizing Carbon- and Water-Footprint Toward Environmentally Sustainable Cloud Computing},
year = {2025},
isbn = {9798400714436},
publisher = {ACM},

booktitle = {Proceedings of the 30th ACM SIGPLAN Annual Symposium on Principles and Practice of Parallel Programming},

series = {PPoPP '25}
}

@article{li2023making,
author = {Li, Pengfei and Yang, Jianyi and Islam, Mohammad A. and Ren, Shaolei},
title = {Making AI Less 'Thirsty'},
year = {2025},
issue_date = {July 2025},
publisher = {ACM},
volume = {68},
number = {7},
issn = {0001-0782},
journal = {Commun. ACM},
}

@article{ZHENG20202208,
title = {Mitigating Curtailment and Carbon Emissions through Load Migration between Data Centers},
journal = {Joule},
volume = {4},
number = {10},
year = {2020},
issn = {2542-4351},
author = {Jiajia Zheng and Andrew A. Chien and Sangwon Suh},
}

@article{plos2022_landuse,
  title={Land-use intensity of electricity production and tomorrow’s energy landscape},
  author={Lovering, Jessica and Swain, Marian and Blomqvist, Linus and Hernandez, Rebecca R},
  journal={PLoS One},
  volume={17},
  number={7},
  year={2022},
  publisher={Public Library of Science},
}

@article{nei,
author = {Aslan, Joshua and Mayers, Kieren and Koomey, Jonathan G. and France, Chris},
title = {Electricity Intensity of Internet Data Transmission: Untangling the Estimates},
journal = {Journal of Industrial Ecology},
volume = {22},
number = {4},
year = {2018}
}

@inproceedings{10.1145/2342356.2342398,
author = {Gao, Peter Xiang and Curtis, Andrew R. and Wong, Bernard and Keshav, Srinivasan},
title = {It's not easy being green},
year = {2012},
isbn = {9781450314190},
publisher = {ACM},
booktitle = {Proceedings of the ACM SIGCOMM 2012 Conference on Applications, Technologies, Architectures, and Protocols for Computer Communication},
series = {SIGCOMM '12}
}

@inproceedings{10.1145/3604930.3605711,
author = {Maji, Diptyaroop and Pfaff, Ben and P R, Vipin and Sreenivasan, Rajagopal and Firoiu, Victor and Iyer, Sreeram and Josephson, Colleen and Pan, Zhelong and Sitaraman, Ramesh K},
title = {Bringing Carbon Awareness to Multi-cloud Application Delivery},
year = {2023},
isbn = {9798400702426},
publisher = {ACM},
booktitle = {Proceedings of the 2nd Workshop on Sustainable Computer Systems},
articleno = {6},
series = {HotCarbon '23}
}

@INPROCEEDINGS {11120561,
author = { Rodrigues, Elvis and Goldverg, Jacob and Kosar, Tevfik },
booktitle = { 2025 IEEE 18th International Conference on Cloud Computing (CLOUD) },
title = {{ Carbon-Aware Temporal Data Transfer Scheduling Across Cloud Datacenters }},
year = {2025},
volume = {},
ISSN = {},
pages = {255-264},
doi = {10.1109/CLOUD67622.2025.00034},
url = {https://doi.ieeecomputersociety.org/10.1109/CLOUD67622.2025.00034},
publisher = {IEEE Computer Society},
address = {Los Alamitos, CA, USA},
month =Jul}

@inproceedings{10.1145/3754598.3754673,
author = {Schweisgut, Dominik and Benoit, Anne and Robert, Yves and Meyerhenke, Henning},
title = {Carbon-Aware Workflow Scheduling with Fixed Mapping and Deadline Constraint},
year = {2025},
isbn = {9798400720741},
publisher = {Association for Computing Machinery},
address = {New York, NY, USA},
url = {https://doi.org/10.1145/3754598.3754673},
doi = {10.1145/3754598.3754673},
booktitle = {Proceedings of the 54th International Conference on Parallel Processing},
pages = {627–637},
numpages = {11},
location = {
},
series = {ICPP '25}
}

@inproceedings{10.1145/3631295.3631396,
author = {Chadha, Mohak and Subramanian, Thandayuthapani and Arima, Eishi and Gerndt, Michael and Schulz, Martin and Abboud, Osama},
title = {GreenCourier: Carbon-Aware Scheduling for Serverless Functions},
year = {2023},
isbn = {9798400704550},
publisher = {ACM},
booktitle = {Proceedings of the 9th International Workshop on Serverless Computing},
numpages = {6},
series = {WoSC '23}
}

@ARTICLE{7345588,
  author={Zhou, Zhi and Liu, Fangming and Zou, Ruolan and Liu, Jiangchuan and Xu, Hong and Jin, Hai},
  journal={IEEE Transactions on Parallel and Distributed Systems}, 
  title={Carbon-Aware Online Control of Geo-Distributed Cloud Services}, 
  year={2016},
  volume={27},
  number={9},
}

@inproceedings{10.1145/3634769.3634812,
author = {Souza, Abel and Jasoria, Shruti and Chakrabarty, Basundhara and Bridgwater, Alexander and Lundberg, Axel and Skogh, Filip and Ali-Eldin, Ahmed and Irwin, David and Shenoy, Prashant},
title = {CASPER: Carbon-Aware Scheduling and Provisioning for Distributed Web Services},
year = {2024},
isbn = {9798400716690},
publisher = {ACM},
booktitle = {Proceedings of the 14th International Green and Sustainable Computing Conference},
series = {IGSC '23}
}

@INPROCEEDINGS{10305816,
  author={Cordingly, Robert and Kaur, Jasleen and Dwivedi, Divyansh and Lloyd, Wes},
  booktitle={2023 IEEE International Conference on Cloud Engineering (IC2E)}, 
  title={Towards Serverless Sky Computing: An Investigation on Global Workload Distribution to Mitigate Carbon Intensity, Network Latency, and Cost}, 
  year={2023},
  volume={},
  number={},
}

@ARTICLE{9770383,
  author={Radovanović, Ana and Koningstein, Ross and Schneider, Ian and Chen, Bokan and Duarte, Alexandre and Roy, Binz and Xiao, Diyue and Haridasan, Maya and Hung, Patrick and Care, Nick and Talukdar, Saurav and Mullen, Eric and Smith, Kendal and Cottman, MariEllen and Cirne, Walfredo},
  journal={IEEE Transactions on Power Systems}, 
  title={Carbon-Aware Computing for Datacenters}, 
  year={2023},
  volume={38},
  number={2},
}

@inproceedings{10.1145/3464298.3493399,
author = {Wiesner, Philipp and Behnke, Ilja and Scheinert, Dominik and Gontarska, Kordian and Thamsen, Lauritz},
title = {Let's wait awhile: how temporal workload shifting can reduce carbon emissions in the cloud},
year = {2021},
isbn = {9781450385343},
publisher = {ACM},
booktitle = {Proceedings of the 22nd International Middleware Conference},
series = {Middleware '21}
}

@INPROCEEDINGS{9975369,
  author={Lee, Sungjae and Hwang, Jaeil and Lee, Kyungyong},
  booktitle={2022 IEEE International Symposium on Workload Characterization (IISWC)}, 
  title={SpotLake: Diverse Spot Instance Dataset Archive Service}, 
  year={2022},
  volume={},
  number={},
  pages={242-255},
  doi={10.1109/IISWC55918.2022.00029}}

@inproceedings{10.1145/3627703.3650079,
author = {Sukprasert, Thanathorn and Souza, Abel and Bashir, Noman and Irwin, David and Shenoy, Prashant},
title = {On the Limitations of Carbon-Aware Temporal and Spatial Workload Shifting in the Cloud},
year = {2024},
isbn = {9798400704376},
publisher = {ACM},
booktitle = {Proceedings of the Nineteenth European Conference on Computer Systems},
series = {EuroSys '24}
}

@inproceedings{10.1145/3447555.3466582,
author = {Lindberg, Julia and Abdennadher, Yasmine and Chen, Jiaqi and Lesieutre, Bernard C. and Roald, Line},
title = {A Guide to Reducing Carbon Emissions through Data Center Geographical Load Shifting},
year = {2021},
isbn = {9781450383332},
publisher = {ACM},
booktitle = {Proceedings of the Twelfth ACM International Conference on Future Energy Systems},
series = {e-Energy '21}
}

@Article{su17146433,
AUTHOR = {Asadov, Nasir and Coroamă, Vlad C. and Franzil, Matteo and Galantino, Stefano and Finkbeiner, Matthias},
TITLE = {Carbon-Aware Spatio-Temporal Workload Shifting in Edge–Cloud Environments: A Review and Novel Algorithm},
JOURNAL = {Sustainability},
VOLUME = {17},
YEAR = {2025},
NUMBER = {14},
ARTICLE-NUMBER = {6433},
ISSN = {2071-1050},
}

@ARTICLE{11250739,
  author={Hall, Sophie and Micheli, Francesco and Belgioioso, Giuseppe and Radovanović, Ana and Dörfler, Florian},
  journal={IEEE Transactions on Power Systems}, 
  title={Carbon-Aware Computing for Data Centers with Probabilistic Performance Guarantees}, 
  year={2025},
  volume={},
  number={},
  pages={1-14},
  doi={10.1109/TPWRS.2025.3630499}}

@INPROCEEDINGS{10749780,
  author={Serenari, Jayden and Sreekumar, Sreekanth and Zhao, Kaiwen and Sarkar, Saurabh and Lee, Stephen},
  booktitle={2024 IEEE International Conference on Cloud Engineering (IC2E)}, 
  title={GreenWhisk: Emission-Aware Computing for Serverless Platform}, 
  year={2024},
  volume={},
  number={},
  pages={44-54},
  doi={10.1109/IC2E61754.2024.00012}}

@article{piontek2023carbon,
  title={Carbon emission-aware job scheduling for Kubernetes deployments},
  author={Piontek, Tobias and Haghshenas, Kawsar and Aiello, Marco},
  journal={Journal of supercomputing},
  volume={80},
  year={2023},
  publisher={Springer},
}

@article{10.1111/jiec.12630,
author = {Aslan, Joshua and Mayers, Kieren and Koomey, Jonathan G. and France, Chris},
title = {Electricity Intensity of Internet Data Transmission: Untangling the Estimates},
journal = {Journal of Industrial Ecology},
volume = {22},
number = {4},
year = {2018}
}

@article{scoutpapaer,
  title={Scout: An experienced guide to find the best cloud configuration},
  author={Hsu, Chin-Jung and Nair, Vivek and Menzies, Tim and Freeh, Vincent W},
  journal={arXiv preprint arXiv:1803.01296},
  year={2018},

}

@INPROCEEDINGS{8276798,
  author={Panneerselvam, John and Liu, Lu and Antonopoulos, Nick},
  booktitle={2017 IEEE International Conference on Internet of Things (iThings) and IEEE Green Computing and Communications (GreenCom) and IEEE Cyber, Physical and Social Computing (CPSCom) and IEEE Smart Data (SmartData)}, 
  title={Characterisation of Hidden Periodicity in Large-Scale Cloud Datacentre Environments}, 
  year={2017},
  volume={},
  number={},

}

@inproceedings{10.5555/3026877.3026887,
author = {Jyothi, Sangeetha Abdu and Curino, Carlo and Menache, Ishai and Narayanamurthy, Shravan Matthur and Tumanov, Alexey and Yaniv, Jonathan and Mavlyutov, Ruslan and Goiri, \'{I}\~{n}igo and Krishnan, Subru and Kulkarni, Janardhan and Rao, Sriram},
title = {Morpheus: towards automated SLOs for enterprise clusters},
year = {2016},
isbn = {9781931971331},
publisher = {USENIX Association},
address = {USA},
booktitle = {Proceedings of the 12th USENIX Conference on Operating Systems Design and Implementation},
series = {OSDI'16},

}

@article{newbold2015global,
  title={Global effects of land use on local terrestrial biodiversity},
  author={Newbold, Tim and Hudson, Lawrence N and Hill, Samantha LL and Contu, Sara and Lysenko, Igor and Senior, Rebecca A and B{\"o}rger, Luca and Bennett, Dominic J and Choimes, Argyrios and Collen, Ben and others},
  journal={Nature},
  volume={520},
  number={7545},
  year={2015},
  publisher={Nature Publishing Group UK London},

}

@inproceedings{azurefaas,
  title={Serverless in the wild: Characterizing and optimizing the serverless workload at a large cloud provider},
  author={Shahrad, Mohammad and Fonseca, Rodrigo and Goiri, Inigo and Chaudhry, Gohar and Batum, Paul and Cooke, Jason and Laureano, Eduardo and Tresness, Colby and Russinovich, Mark and Bianchini, Ricardo},
  booktitle={2020 USENIX annual technical conference (USENIX ATC 20)},
  year={2020},

}

@inproceedings{alfadda2017hour,
  title={Hour-ahead solar PV power forecasting using SVR based approach},
  author={Alfadda, Abdullah and Adhikari, Rajendra and Kuzlu, Murat and Rahman, Saifur},
  booktitle={2017 IEEE Power \& Energy Society Innovative Smart Grid Technologies Conference (ISGT)},
  year={2017},
  organization={IEEE}, 

}

@article{okumus2016current,
  title={Current status of wind energy forecasting and a hybrid method for hourly predictions},
  author={Okumus, Inci and Dinler, Ali},
  journal={Energy Conversion and Management},
  volume={123},
  year={2016},
  publisher={Elsevier},
}

@article{zayas2025new,
  title={A new deep learning-based approach for predicting the geothermal heat pump’s thermal power of a real bioclimatic house},
  author={Zayas-Gato, Francisco and D{\'\i}az-Longueira, Antonio and Arcano-Bea, Paula and Michelena, {\'A}lvaro and Calvo-Rolle, Jose Luis and Jove, Esteban},
  journal={Applied Intelligence},
  volume={55},
  number={6},
  year={2025},
  publisher={Springer}, 
}

@article{bilgili2022one,
  title={One-day ahead forecasting of energy production from run-of-river hydroelectric power plants with a deep learning approach},
  author={Bilgili, Mehmet and Keiyinci, Sinan and Ekinci, Firat},
  journal={Scientia Iranica},
  volume={29},
  number={4},
  year={2022},
  publisher={Sharif University of Technology}
}

@article{loco24,
      title={Environmentally-Conscious Cloud Orchestration Considering Geo-Distributed Data Centers}, 
      author={Giulio Attenni and Novella Bartolini},
      year={2025},
      journal={Post-Proceedings of the 1st International Workshop on Low Carbon Computing, LOCO '24},
      publisher={arXiv-based Proceedings}
}

@article{belkhir2018assessing,
  title={Assessing ICT global emissions footprint: Trends to 2040 \& recommendations},
  author={Belkhir, Lotfi and Elmeligi, Ahmed},
  journal={Journal of cleaner production},
  volume={177},
  year={2018},
  publisher={Elsevier}, 

}

@article{FREITAG2021100340,
title = {The real climate and transformative impact of ICT: A critique of estimates, trends, and regulations},
journal = {Patterns},
volume = {2},
number = {9},
year = {2021},
issn = {2666-3899},
author = {Charlotte Freitag and Mike Berners-Lee and Kelly Widdicks and Bran Knowles and Gordon S. Blair and Adrian Friday},
}

@article{10.1145/3626788,
author = {Hanafy, Walid A. and Liang, Qianlin and Bashir, Noman and Irwin, David and Shenoy, Prashant},
title = {CarbonScaler: Leveraging Cloud Workload Elasticity for Optimizing Carbon-Efficiency},
year = {2023},
issue_date = {December 2023},
publisher = {ACM},
volume = {7},
number = {3},
journal = {Proc. ACM Meas. Anal. Comput. Syst.},
month = dec,
articleno = {57},

}

@inproceedings{10.1145/3575813.3595197,
author = {Lin, Liuzixuan and Chien, Andrew A},
title = {Adapting Datacenter Capacity for Greener Datacenters and Grid},
year = {2023},
isbn = {9798400700323},
publisher = {ACM},
booktitle = {Proceedings of the 14th ACM International Conference on Future Energy Systems},
series = {e-Energy '23}
}

@ARTICLE{TCC.2015.2453982,
  author={Islam, Mohammad A. and Ren, Shaolei and Quan, Gang and Shakir, Muhammad Z. and Vasilakos, Athanasios V.},
  journal={IEEE Transactions on Cloud Computing}, 
  title={Water-Constrained Geographic Load Balancing in Data Centers}, 
  year={2015},
  volume={5},
  number={2},

}

@misc{owid_safest_sources,
  title={What are the safest and cleanest sources of energy?},
  author={{Our World in Data}},
  year={2020},
  url={https://ourworldindata.org/safest-sources-of-energy}, 
  note={accessed: 2026-02-20}
}

@article{10.14778/2556549.2556553,
author = {Popescu, Adrian Daniel and Balmin, Andrey and Ercegovac, Vuk and Ailamaki, Anastasia},
title = {PREDIcT: towards predicting the runtime of large scale iterative analytics},
year = {2013},
issue_date = {September 2013},
publisher = {VLDB Endowment},
volume = {6},
number = {14},
issn = {2150-8097},
journal = {Proc. VLDB Endow.},
numpages = {12}
}

@INPROCEEDINGS{9610321,
  author={Will, Jonathan and Thamsen, Lauritz and Scheinert, Dominik and Bader, Jonathan and Kao, Odej},
  booktitle={2021 IEEE International Conference on Cloud Engineering (IC2E)}, 
  title={C3O: Collaborative Cluster Configuration Optimization for Distributed Data Processing in Public Clouds}, 
  year={2021},
  volume={},
  number={},}

@inproceedings{10.1145/3514221.3517892,
author = {Al-Sayeh, Hani and Memishi, Bunjamin and Jibril, Muhammad Attahir and Paradies, Marcus and Sattler, Kai-Uwe},
title = {Juggler: Autonomous Cost Optimization and Performance Prediction of Big Data Applications},
year = {2022},
isbn = {9781450392495},
publisher = {Association for Computing Machinery},
booktitle = {Proceedings of the 2022 International Conference on Management of Data},
series = {SIGMOD '22}
}

@inproceedings {194946,
author = {Shivaram Venkataraman and Zongheng Yang and Michael Franklin and Benjamin Recht and Ion Stoica},
title = {Ernest: Efficient Performance Prediction for {Large-Scale} Advanced Analytics},
booktitle = {13th USENIX Symposium on Networked Systems Design and Implementation (NSDI 16)},
year = {2016},
isbn = {978-1-931971-29-4},
publisher = {USENIX Association},
}

@inproceedings{10.1145/3632775.3639589,
author = {Wiesner, Philipp and Khalili, Ramin and Grinwald, Dennis and Agrawal, Pratik and Thamsen, Lauritz and Kao, Odej},
title = {FedZero: Leveraging Renewable Excess Energy in Federated Learning},
year = {2024},
publisher = {ACM},
booktitle = {Proceedings of the 15th ACM International Conference on Future and Sustainable Energy Systems},
pages = {373–385},
series = {e-Energy '24}
}

@inproceedings{10.1007/978-3-031-12597-3_14,
author = {Wiesner, Philipp and Scheinert, Dominik and Wittkopp, Thorsten and Thamsen, Lauritz and Kao, Odej},
title = {Cucumber: Renewable-Aware Admission Control for Delay-Tolerant Cloud and Edge Workloads},
year = {2022},
publisher = {Springer},
booktitle = {Euro-Par 2022: Parallel Processing: 28th International Conference on Parallel and Distributed Computing},
pages = {218–232},
}

@inproceedings{11044837,
  author={West, Kathleen and Lehmann, Fabian and Bountris, Vasilis and Leser, Ulf and Elkhatib, Yehia and Thamsen, Lauritz},
  booktitle={2025 IEEE 25th International Symposium on Cluster, Cloud and Internet Computing}, 
  title={Exploring the Potential of Carbon-Aware Execution for Scientific Workflows}, 
  year={2025},
  publisher={IEEE},
  pages={325-328},
  series = {CCGrid '25}
}

@techreport{iea_dc2030,
    author = {},
    title = {Energy and AI},
    institution = {IEA},
    year = {2025}, 
    url={https://www.iea.org/reports/energy-and-ai},
    note= {Accessed: 2026-02-20}
}

@article{pue,
  title={PUE: a comprehensive examination of the metric},
  author={Avelar, Victor and Azevedo, Dan and French, Alan and Power, Emerson Network},
  journal={White paper},
  volume={49},
  pages={52},
  year={2012}, 
  url={https://www.thegreengrid.org/en/resources/library-and-tools/20-PUE%3A-A-Comprehensive-Examination-of-the-Metric}, 
  note= {Accessed: 2026-02-20}
}

@article{wue,
  title={Water usage effectiveness (WUE): A green grid datacenter sustainability metric},
  author={Azevedo, Dan and Belady, Symantec Christian and Pouchet, Jack},
  journal={The Green Grid},
  volume={32},
  pages={16},
  year={2011},
  url={https://airatwork.com/wp-content/uploads/The-Green-Grid-White-Paper-35-WUE-Usage-Guidelines.pdf},
  note= {Accessed: 2026-02-20}
}

@techreport{lue,
    author = {Kartik Sumrani},
    title = {Decarbonization: Towards Sustainable Data Centers},
    institution = {Dell Technologies},
    year = {2023}, 
    url={https://learning.dell.com/content/dam/dell-emc/documents/en-us/2023KS_Sumrani-Decarbonization_To_Sustainable_Data_Centers.pdf},
    note= {Accessed: 2026-02-20}
}
\sloppy
\appendix
\section{\review{Research Materials — Full Source Documentation}}
\label{online_res}

\review{
\subsection{Workload Traces}
\label{app:workload}
\begin{itemize}
    \item \textbf{Azure Functions (FaaS)}: 2019 Azure Functions trace, providing per-minute invocation counts and daily execution-time percentiles (0th, 1st, 25th, 50th, 75th, 99th, 100th) for each function. Traces can be found in the Azure official GitHub profile. The specific dataset is available at \url{https://github.com/Azure/AzurePublicDataset/blob/master/AzureFunctionsDataset2019.md}
    \item \textbf{AWS Big Data (Spark)}: job configuration and runtime traces from the \emph{scout} tool for cloud configuration optimization. Available at the GitHub repository "scout: A Tool for Cloud Configuration Optimization" - \url{https://github.com/oxhead/scout}

\end{itemize}

}
\review{
\subsection{Data Center Efficiency and Land Occupation}
\label{app:dc_characteristics}
Table~\ref{tab:dc_sources} reports the data sources of Data Center Efficiency metrics such as PUE and WUE, and sources of Land Occupation data.
}

{
\begin{table*}[ht!]
\centering
\small
\begin{tabular}{p{1.5cm}p{2cm}p{0.35\textwidth}p{0.35\textwidth}}
\hline
\textbf{Provider} & \textbf{Region} & \textbf{Sources PUE/WUE} & \textbf{Sources Land} \\
\hline
Azure & swedencentral & 
\url{https://datacenters.microsoft.com/wp-content/uploads/2024/08/Sweden%20%28Sweden%20Central%29.pdf}&
Article: "Two municipalities in the Stockholm region sell land to Microsoft", published by Invest Stockholm
\url{https://www.mynewsdesk.com/investstockholm/pressreleases/two-municipalities-in-the-stockholm-region-sell-land-to-microsoft-2813124} \\
\hline
Azure & southcentralus & 
Microsoft Sustainability Factsheets \url{https://datacenters.microsoft.com/wp-content/uploads/2024/08/Texas%20%28South%20Central%20US%29.pdf}&
Datacenters.com, "Microsoft Azure: South Central US-Texas" \url{https://www.datacenters.com/microsoft-azure-south-central-us-texas} \\
\hline
Azure & centralus & 
Microsoft Sustainability Factsheets \url{https://datacenters.microsoft.com/wp-content/uploads/2024/08/Iowa%20%28Central%20US%29.pdf}&
Datacenters.com, "Microsoft Azure: Central US-Iowa"
\url{https://www.datacenters.com/microsoft-azure-des-moines} \\
\hline
Azure & eastus2 & 
Microsoft Sustainability Factsheets from Internet Archive's Wayback Machine Snapshots \url{https://web.archive.org/web/20251113020848/https://datacenters.microsoft.com/globe/pdfs/sustainability/factsheets/Virginia%20(East%20US%20&%20East%20US%202).pdf}&
Datacenters.com, "Microsoft Azure: East US-Virginia"
\url{https://www.datacenters.com/microsoft-azure-east-us-virginia} \\
\hline
AWS & all 5 regions & 
Microsoft Sustainability Factsheets \url{https://sustainability.aboutamazon.com/products-services/aws-cloud}&
FORM 10-K, land estimated via AZ-count share of total
\url{https://s2.q4cdn.com/299287126/files/doc_financials/2025/ar/Amazon-2024-Annual-Report.pdf}  \\
\hline
\end{tabular}
\caption{Data center efficiency (PUE/WUE) and land occupation sources, by region.}
\label{tab:dc_sources}
\end{table*}
}

\review{
\subsection{VM Instance Specifications (AWS)}
\label{app:vm}
Instances memory and cpu can be found at 
\url{https://aws.amazon.com/ec2/instance-types}, and bandwidth values at
\url{https://docs.aws.amazon.com/ec2/latest/instancetypes/pg.html}.
}

\subsection{Migration Monetary Costs (AWS)}
\label{app:migrcosts}
AWS charges for data transfer out from an Amazon EC2 based on the destination region. A table summarizing the costs is available at:
\url{https://aws.amazon.com/ec2/pricing/on-demand/#Data_Transfer}. Charging rates used in our evaluation are also reported in Table~\ref{tab:aws_transf_charge}

\begin{table}[]
    \centering
    \begin{tabular}{c|c}
       Destination Region  &  USD/GB Transferred\\\hline
       eu-central-1  & 0.05 \\\hline
       eu-west-2  & 0.02 \\\hline
       eu-north-1  &  0.02\\\hline
       us-east-1  & 0.01 \\\hline
       us-west-1  & 0.02 \\\hline
       
    \end{tabular}
    \caption{AWS charges for data transferring}
    \label{tab:aws_transf_charge}
\end{table}

\subsection{Electricity Grid Mix Data}
Timeseries of power grid mix.
\begin{itemize}
    \item \emph{Sweden}: ENTSOE Transparency Platform \url{https://transparency.entsoe.eu/}.
    \item \emph{United Kingdom}: National Electricity System Operator API \url{https://carbon-intensity.github.io/api-definitions/}.
    \item \emph{Germany}: SMARD - Download market data \url{https://www.smard.de/en/downloadcenter/download-market-data/}.
    \item \emph{US}: EIA - Wholesale Electricity Market Data. Table~\ref{tab:iea_us_data} shows the URLs of the regions that we used in our work.
    
\end{itemize}

\begin{table}[]
    \centering
    \begin{tabular}{c|p{5cm}}
       Region  &  URL\\\hline
       ERCOT  & \url{https://www.eia.gov/electricity/wholesalemarkets/data.php?rto=ercot} \\\hline
       MISO  & \url{https://www.eia.gov/electricity/wholesalemarkets/data.php?rto=miso} \\\hline
       PJM  & \url{https://www.eia.gov/electricity/wholesalemarkets/data.php?rto=pjm} \\\hline
       CAISO  & \url{https://www.eia.gov/electricity/wholesalemarkets/data.php?rto=caiso} \\\hline
       
    \end{tabular}
    \caption{US electric grid regions' URLs}
    \label{tab:iea_us_data}
\end{table}

\review{
\subsection{Impact Coefficients (Carbon, Water, Land)}
\label{app:coefficients}
Source-level documentation for Table~\ref{tab:intensities}:
\begin{itemize}
    \item \textbf{Carbon intensity}: IPCC AR5 Annex III, Table A.III.2 (median lifecycle emissions)\cite{ipcc2014_ar5_annexIII}; 
    oil coefficient from Our World in Data, \url{https://ourworldindata.org/safest-sources-of-energy}.
    \item \textbf{Water intensity (EWIF)}: NREL technical report on energy-related water consumption, Tables 1–2 \cite{nrel2011_water}.
    \item \textbf{Land intensity (ELIF)}: PLOS land-use survey (oil assumed comparable to gas, per survey authors) \cite{plos2022_landuse}.
\end{itemize}
}
\newpage

\end{document}